\documentclass[11pt]{article}
\usepackage{times}
\usepackage{float}
\input epsf.sty

\newcommand{\comment}[1]{}

\newcommand{\qed}{\nobreak \ifvmode \relax \else
      \ifdim\lastskip<1.5em \hskip-\lastskip
      \hskip1.5em plus0em minus0.5em \fi \nobreak
      \vrule height0.75em width0.5em depth0.25em\fi}

\def \lket {|}
\def \rket {\rangle}
\def \lbra {\langle}

\def \rbra {|}
\def \H {{\cal H}}
\def \F {{\cal F}}
\def \ctg{\cot}
\newcommand{\ket}[1]{\lket #1\rket}
\newcommand{\bra}[1]{\lbra #1\rbra}
\newtheorem{Definition}{Definition}
\newtheorem{Theorem}{Theorem}
\newtheorem{Lemma}{Lemma}
\newtheorem{Claim}{Claim}
\newcommand{\proof}{\noindent {\bf Proof: }}
\floatstyle{boxed}
\newfloat{Algorithm}{bt}{lop}
\floatname{Algorithm}{Algorithm}
\begin{document}
\title{Quantum walk algorithm for element distinctness}
\date{}
\author{Andris Ambainis\thanks{
Department of Combinatorics and Optimization, Faculty of Mathematics,
University of Waterloo, 200 University Avenue West, Waterloo, ON N2L 2T2, Canada,
e-mail: {\tt ambainis@math.uwaterloo.ca}. Parts of this research done at 
University of Latvia, University of California, Berkeley 
and Institute for Advanced Study, Princeton.
Supported by Latvia Science Council Grant 01.0354 (at University of Latvia),
DARPA and Air Force Laboratory, Air Force Materiel Command, USAF, under
agreement number F30602-01-2-0524 (at UC Berkeley), NSF Grant DMS-0111298 (at IAS), 
NSERC, ARDA, IQC University Professorship and CIAR (at University of 
Waterloo).}}
\maketitle

\begin{abstract}
We use quantum walks to construct a new quantum algorithm
for element distinctness and its generalization.
For element distinctness (the problem of finding two equal
items among $N$ given items), we get an $O(N^{2/3})$
query quantum algorithm.
This improves the previous $O(N^{3/4})$ 
quantum algorithm
of Buhrman et al.   \cite{Distinctness} and matches the lower
bound by \cite{Aaronson}.
We also give an $O(N^{k/(k+1)})$ query
quantum algorithm for the generalization of element 
distinctness in which we have to find $k$ equal items among $N$
items.
\end{abstract}

\section{Introduction}

Element distinctness is the following problem.

{\bf Element Distinctness.}
Given numbers $x_1, \ldots, x_N\in[M]$, are they all distinct?

It has been extensively studied both in classical and quantum
computing. Classically, the best way to solve element distinctness 
is by sorting which requires $\Omega(N)$ queries.
In quantum setting, Buhrman et al.   \cite{Distinctness} have 
constructed a quantum algorithm that uses $O(N^{3/4})$ queries.
Aaronson and Shi \cite{Aaronson} have shown that any quantum algorithm requires
at least $\Omega(N^{2/3})$ quantum queries. 

In this paper, we give a new quantum algorithm that solves element distinctness
with $O(N^{2/3})$ queries to $x_1, \ldots, x_N$.
This matches the lower bound of \cite{Aaronson,ASmallRange}.

Our algorithm uses a combination of several ideas:
quantum search on graphs \cite{AA} and quantum walks \cite{Kempe}.
While each of those ideas has been used before,
the present combination is new.

We first reduce element distinctness to searching a
certain graph with vertices $S\subseteq\{1, \ldots, N\}$ as 
vertices. The goal of the search is to find a marked
vertex. Both examining the current vertex and moving to
a neighboring vertex cost one time step. (This contrasts
with the usual quantum search \cite{Grover}, 
where only examining the current vertex costs one time step.)

We then search this graph by quantum random walk.
We start in a uniform superposition 
over all vertices of a graph and perform a quantum random walk
with one transition rule for unmarked vertices of the graph and 
another transition rule for marked vertices of the graph.
The result is that the amplitude gathers in the marked vertices
and, after $O(N^{2/3})$ steps, the probability of measuring 
the marked state is a constant.

We also give several extensions of our algorithm.
If we have to find whether $x_1$, $\ldots$, $x_N$ contain
$k$ numbers that are equal: $x_{i_1}=\ldots=x_{i_k}$,
we get a quantum algorithm with $O(N^{k/(k+1)})$
queries for any constant\footnote{The big-O constant depends on $k$. 
For non-constant $k$, we can show that the number of queries
is $O(k^2 N^{k/(k+1)})$. The proof of that is mostly technical
and is omitted in this version. } $k$. 

If the quantum algorithm is restricted to storing
$r$ numbers, $r\leq N^{2/3}$, then we have an algorithm which solves
element distinctness with $O(N/\sqrt{r})$ queries 
which is quadratically better than the classical
$O(N^2/r)$ query algorithm. Previously, such quantum algorithm was
known only for $r\leq \sqrt{N}$ \cite{Distinctness}.
For the problem of finding $k$ equal numbers, we
get an algorithm that uses $O(\frac{N^{k/2}}{r^{(k-1)/2}})$ queries
and stores $r$ numbers, for $r\leq N^{(k-1)/k}$.

For the analysis of our algorithm, we develop a generalization
of Grover's algorithm (Lemma \ref{lem:gengrover}) 
which might be of independent interest.


\subsection{Related work}
{\bf Classical element distinctness.}
Element distinctness has been extensively studied classically.
It can be solved with $O(N)$ queries and $O(N\log N)$ time
by querying all the elements and sorting them.
Then, any two equal elements must be next one to another
in the sorted order and can be found by going through
the sorted list.

In the usual query model (where one query gives one value of $x_i$),
it is easy to see that $\Omega(N)$ queries are also necessary.
Classical lower bounds have also been shown for more general
models (e.g. \cite{Grig}).

The algorithm described above requires $\Omega(N)$ space to store
all of $x_1, \ldots, x_N$.
If we are restricted to space $S<N$, the running time increases.
The straightforward algorithm needs $O(\frac{N^2}{S})$ queries.
Yao \cite{Yao} has shown that, for the model of comparison-based
branching programs, this is essentially optimal.
Namely, any space-$S$ algorithm needs time $T=\Omega(\frac{N^{2-o(1)}}{S})$.
For more general models, lower bounds on algorithms with restricted 
space $S$ is an object of ongoing research \cite{BSSV}. 

{\bf Related problems in quantum computing.}
In {\em collision problem},  
we are given a 2-1 function $f$ and have to find $x, y$
such that $f(x)=f(y)$.
As shown by Brassard, H\o yer and Tapp \cite{BHT},
collision problem can be solved in $O(N^{1/3})$ quantum 
steps instead of
$\Theta(N^{1/2})$ steps classically.
$\Omega(N^{1/3})$ 
is also a quantum lower bound \cite{Aaronson,Kutin}. 

If element distinctness
can be solved with $M$ queries, then collision
problem can be solved with $O(\sqrt{M})$ queries. 
(This connection is credited to Andrew Yao in \cite{Aaronson}.)
Thus, a quantum algorithm for element distinctness implies
a quantum algorithm for collision but not the other
way around. 

{\bf Quantum search on graphs.}
The idea of quantum search on graphs was proposed by 
Aaronson and Ambainis \cite{AA}
for finding a marked item on a $d$-dimensional grid 
(problem first considered by Benioff \cite{Benioff}) 
and other graphs with good expansion properties.
Our work has a similar flavor but uses
completely different methods to search the graph
(quantum walk instead of ``divide-and-conquer'').

{\bf Quantum walks.}
There has been considerable amount of research
on quantum walks (surveyed in \cite{Kempe})
and their applications (surveyed in \cite{ASurvey1}).
Applications of walks \cite{ASurvey1} mostly fall into two classes.
The first class is exponentially faster hitting times \cite{CF+,CC+,KempeHit}.
\comment{In these applications, quantum walks are used to find a certain vertex exponentially
faster than classically. \comment{(The first results \cite{FG,CF+,KempeHit} were quantum walk 
algorithms that are exponentially faster than classical algorithms based on classical random walks.
The most recent result \cite{CC+} is a quantum walk algorithm which is exponentially
faster than any classical algorithm.)}
Our result is quite different from those, both in the problem
that we consider and the way how we solve.
While polynomial speedup achieved in our paper 
is smaller than exponential speedup in \cite{CC+},
our result has the advantage that we solve a natural problem
which has been widely studied in both classical
and quantum computing. 
}
The second class is quantum walk search algorithms \cite{Shenvi,CG,AKRS}.

Our algorithm is most closely related to the second class.
In this direction, Shenvi et al. \cite{Shenvi} have 
constructed a counterpart of Grover's search \cite{Grover} 
based on quantum walk on the hypercube. 
Childs and Goldstone \cite{CG,CG04} and 
Ambainis et al. \cite{AKRS} have used quantum walk to produce search
algorithms on $d$-dimensional lattices ($d\geq 2$)
which is faster than the naive application of Grover's search.
This direction is quite closely related to our work.
The algorithms by \cite{Shenvi,CG,AKRS} and current paper
solve different problems but all have similar structure.

{\bf Recent developments.}
After the work described in this paper, the results and ideas 
from this paper have been used to construct several other
quantum algorithms. Magniez et al.  \cite{MSS} have used 
our element distinctness algorithm to give an $O(n^{1.3})$ query
quantum algorithm for finding triangles in a graph.
Ambainis et al. \cite{AKRS} have used ideas from the current paper to 
construct a faster algorithm for search on 2-dimensional grid.
Childs and Eisenberg \cite{CE} have given a different analysis
of our algorithm.

Szegedy \cite{Szegedy} has generalized our results on
quantum walk for element distinctness to
an arbitrary graph with a large eigenvalue gap and cast
them into the language of Markov chains.
His main result is
that, for a class of Markov chains, quantum walk algorithms are 
quadratically faster than the corresponding classical algorithm.
An advantage of Szegedy's approach is that it can simultaneously handle 
any number of solutions (unlike in the present paper which has separate
algorithms for single solution case (algorithm \ref{MainAlg}) and 
multiple-solution case (algorithm \ref{MultAlg})).

Buhrman and Spalek \cite{BS} have used Szegedy's result to
construct an $O(n^{5/3})$ quantum algorithm for verifying if
a product of two $n\times n$ matrices $A$ and $B$ is equal to
a third matrix $C$.

\section{Preliminaries}

\subsection{Quantum query algorithms}

Let $[N]$ denote $\{1, \ldots, N\}$.
We consider 

{\bf Element Distinctness.}
Given numbers $x_1, \ldots, x_N\in[M]$, are there $i, j\in[N]$, $i\neq j$ such
that $x_i=x_j$?

Element distinctness is a particular case of 

{\bf Element $k$-distinctness.}
Given numbers $x_1, \ldots, x_N\in[M]$, are there $k$ distinct indices
$i_1, \ldots, i_k\in [N]$ such that
$x_{i_1}=x_{i_2}=\ldots=x_{i_k}$?

We call such $k$ indices $i_1, \ldots, i_k$ a $k${\em -collision}.

Our model is the quantum query model (for surveys on query model, 
see \cite{ASurvey,BWSurvey}).
In this model, our goal is to compute a function $f(x_1, \ldots, x_N)$.
For example, $k$-distinctness is viewed as the function
$f(x_1, \ldots, x_N)$ which is 1 if there exists a $k$-collision 
consisting of $i_1, \ldots, i_k\in [N]$ and 0 otherwise.

The input variables $x_i$ 
can be accessed by queries to an oracle $X$
and the complexity of $f$ is the number of queries needed to compute $f$.
A quantum computation with $T$ queries
is just a sequence of unitary transformations
\[ U_0\rightarrow O\rightarrow U_1\rightarrow O\rightarrow\ldots
\rightarrow U_{T-1}\rightarrow O\rightarrow U_T.\]

$U_j$'s can be arbitrary unitary transformations that do not depend
on the input bits $x_1, \ldots, x_N$. $O$ are query (oracle) transformations.
To define $O$, we represent basis states as $|i, a, z\rangle$ where
$i$ consists of $\lceil \log N\rceil$ bits, $a$ consists of $\lceil \log M\rceil$ quantum bits
and $z$ consists of all other bits. Then, $O$ maps
$\ket{i, a, z}$ to $\ket{i, (a+x_i) \bmod M, z}$.

In our algorithm, we use queries in two situations.
The first situation is when $a=\ket{0}$. Then, the state
before the query is some superposition $\sum_{i,z} \alpha_{i, z} \ket{i, 0, z}$
and the state after the query is the same superposition with
the information about $x_i$: $\sum_{i,z} \alpha_{i, z} \ket{i, x_i, z}$.
The second situation is when the state before the query is 
$\sum_{i,z} \alpha_{i, z} \ket{i, -x_i \bmod M, z}$ 
with the information about $x_i$ from a previous query.
Then, applying the query transformation makes the state
$\sum_{i,z} \alpha_{i, z} \ket{i, 0, z}$, erasing the information about $x_i$.
This can be used to erase the information about $x_i$ from 
$\sum_{i,z} \alpha_{i, z} \ket{i, x_i, z}$.
We first perform a unitary that maps $\ket{x_i}\rightarrow \ket{-x_i \bmod M}$,
obtaining the state $\sum_{i,z} \alpha_{i, z} \ket{i, -x_i \bmod M, z}$
and then apply the query transformation.

The computation starts with a state $|0\rangle$.
Then, we apply $U_0$, $O$, $\ldots$, $O$,
$U_T$ and measure the final state.
The result of the computation is the rightmost bit of
the state obtained by the measurement.

We say that the quantum computation computes
$f$ with bounded error if, for every $x=(x_1, \ldots, x_N)$,
the probability that the rightmost bit 
of $U_T O_x U_{T-1} \ldots O_x U_0\ket{0}$ 
equals  $f(x_1, \ldots, x_N)$ is at
least $1-\epsilon$ for some fixed $\epsilon<1/2$.

To simplify the exposition, we  occasionally describe a quantum computation
as a classical algorithm with several quantum subroutines
of the form $U_t O_x U_{t-1} \ldots O_x U_0\ket{0}$. 
Any such classical 
algorithm with quantum subroutines
can be transformed into an equivalent sequence 
$U_T O_x U_{T-1} \ldots O_x U_0\ket{0}$
with the number of queries being equal to the 
number of queries in the classical algorithm 
plus the sum of numbers of queries in all quantum subroutines. 

{\bf Comparison oracle.}
In a different version of query model, we are only allowed comparison queries.
In a comparison query, we give two indices $i, j$ to the oracle.
The oracle answers whether $x_i<x_j$ or $x_i\geq x_j$.
In the quantum model, we can query the comparison oracle with
a superposition $\sum_{i, j, z} a_{i, j, z}\ket{i, j, z}$,
where $i, j$ are the indices being queried and $z$ is the rest 
of quantum state. The oracle then performs a unitary
transformation $\ket{i, j, z}\rightarrow -\ket{i, j, z}$ for
all $i, j, z$ such that $x_i<x_j$ and 
$\ket{i, j, z}\rightarrow \ket{i, j, z}$ for  
all $i, j, z$ such that $x_i\geq x_j$.
In section \ref{sec:misc}, we show that our algorithms 
can be adapted to this model
with a logarithmic increase in the number of queries.

\subsection{$d$-wise independence}

To make our algorithms efficient in terms of running time and,
in the case of multiple-solution algorithm in section \ref{sec:multiple}, 
also space, we use $d$-wise independent functions.
A reader who is only interested in the query complexity
of the algorithms may skip this subsection.

\begin{Definition}
Let ${\cal F}$ be a family of functions $f:[N]\rightarrow \{0, 1\}$.
${\cal F}$ is $d$-wise independent if, for all $d$-tuples of 
pairwise distinct $i_1, \ldots, i_d\in [N]$ and all 
$c_1, \ldots, c_d\in\{0, 1\}$,
\[ Pr[f(i_1)=c_1, f(i_2)=c_2, \ldots, f(i_d)=c_d] = \frac{1}{2^d} .\]
\end{Definition}

\begin{Theorem}
\label{thm:kwise1}
\cite{ABI}
There exists a $d$-wise independent family 
$\F=\{f_j|j\in[R]\}$ of functions
$f_j:[N]\rightarrow\{0, 1\}$ such that:
\begin{enumerate}
\item
$R=O(N^{\lceil d/2\rceil})$;
\item
$f_j(i)$ is computable in $O(d \log^2 N)$ time, given $j$ and $i$.
\end{enumerate} 
\end{Theorem}

We will also use families of permutations with a similar properties.
It is not known how to construct small $d$-wise independent
families of permutations. There are, however, constructions
of approximately $d$-wise independent families of permutations.

\begin{Definition}
\label{def:approx-k}
Let ${\cal F}$ be a family of permutations on $f:[n]\rightarrow [n]$.
${\cal F}$ is $\epsilon$-approximately $d$-wise independent if, for all
$d$-tuples of pairwise distinct $i_1, \ldots, i_d\in [n]$ and 
pairwise distinct $j_1, \ldots, j_d\in [n]$, 
\[ Pr[f(i_1)=j_1, f(i_2)=j_2, \ldots, f(i_d)=j_d] \in \left[
\frac{1-\epsilon}{n(n-1)\ldots (n-d+1)}, 
\frac{1+\epsilon}{n(n-1)\ldots (n-d+1)} 
\right] .\]
\end{Definition}

\begin{Theorem}
\label{thm:kwise2}
\cite{Itoh}
Let $n$ be an even power of a prime number.
For any $d\leq n$, $\epsilon>0$, 
there exists an $\epsilon$-approximate $d$-wise independent 
family $\F=\{\pi_j|j\in[R]\}$ of permutations
$\pi_j:[n]\rightarrow [n]$ such that:
\begin{enumerate}
\item
$R=O((n^{d^2}/\epsilon^d)^{3+o(1)})$;
\item
$\pi_j(i)$ is computable in $O(d \log^2 n)$ time, given $j$ and $i$.
\end{enumerate} 
\end{Theorem}

\section{Results and algorithms}

Our main results are

\begin{Theorem}
\label{thm:1}
Element $k$-distinctness can be solved by a quantum algorithm 
with $O(N^{k/(k+1)})$ queries.
In particular, element distinctness can be solved by 
a quantum algorithm with $O(N^{2/3})$ queries.
\end{Theorem}

\begin{Theorem}
\label{thm:2}
Let $r\geq k$, $r=o(N)$.
There is a quantum algorithm that solves element distinctness
with $O(\max(\frac{N}{\sqrt{r}}, r))$ queries and 
and $k$-distinctness with
$O(\max(\frac{N^{k/2}}{r^{(k-1)/2}}, r))$ queries,
using $O(r (\log M+\log N))$ qubits of memory.
\end{Theorem}

Theorem \ref{thm:1} follows from Theorem \ref{thm:2} by setting
$r=\lfloor N^{2/3} \rfloor$ for element distinctness
and $r=\lfloor N^{k/(k+1)}\rfloor$ for $k$-distinctness.
(These values minimize the expressions for the number of
queries in Theorem \ref{thm:2}.)


Next, we present Algorithms \ref{MainAlg} 
which solves element distinctness if we have 
a promise that $x_1, \ldots, x_N$ are either all distinct or 
there is exactly one pair $i, j$, $i\neq j$, $x_i=x_j$
(and $k$-distinctness if we have a promise that there is
at most one set of $k$ indices $i_1, \ldots, i_k$ such that
$x_{i_1}=x_{i_2}=\ldots=x_{i_k}$).
The proof of correctness of algorithm \ref{MainAlg} 
is given in section \ref{sec:MainAn}.
After that, in section \ref{sec:multiple}, we present
Algorithm \ref{MultAlg} which solves the general case, using
Algorithm \ref{MainAlg} as a subroutine.

\subsection{Main ideas}

We start with an informal description of main ideas.
For simplicity, we restrict to element distinctness
and postpone the more general $k$-distinctness 
till the end of this subsection. 

Let $r=N^{2/3}$. 
We define a graph $G$ with ${N \choose r}+{N \choose r+1}$ vertices.
The vertices $v_S$ correspond to sets $S\subseteq [N]$ of size $r$ and $r+1$.
Two vertices $v_S$ and $v_T$ are connected by an edge if 
$T=S\cup\{i \}$ for some $i\in[N]$.
\comment{If we are in a vertex $v_S$, we know $S$ and $x_k$ for all $k\in S$
(i.e. our state contains $\ket{S}$ in one register and
$\otimes_{k\in S} \ket{x_k}$ in another register).}
A vertex is marked if $S$ contains $i, j$, $x_i=x_j$.

Element distinctness reduces to finding a marked vertex in this graph.
If we find a marked vertex $v_S$, then we know that $x_i=x_j$ for some $i, j\in S$, 
i.e. $x_1, \ldots, x_N$ are not all distinct.

The naive way to find a marked vertex 
would be to use 
Grover's quantum search algorithm \cite{Grover,BHMT}.
If $\epsilon$ fraction of vertices are marked,
then Grover's search finds a marked vertex after 
$O(\frac{1}{\sqrt{\epsilon}})$ vertices.
Assume that there exists a single pair
$i, j\in[N]$ such that $i\neq j$, $x_i=x_j$.
For a random $S$, $|S|=N^{2/3}$, the probability of $v_S$ being marked
is 
\[ Pr [i\in S;j\in S]= Pr[i\in S] Pr[j\in S|i\in S] = 
 \frac{N^{2/3}}{N} \frac{N^{2/3}-1}{N-1} = (1-o(1)) \frac{1}{N^{2/3}}.\]
Thus, a quantum algorithm can find a marked vertex by 
examining $O(\frac{1}{\sqrt{\epsilon}})=O(N^{1/3})$ vertices.
However, to find out if a vertex is marked, the algorithm needs
to query $N^{2/3}$ items $x_i$, $i\in S$.
This makes the total query complexity $O(N^{1/3} N^{2/3})=O(N)$,
giving no speedup compared to the classical algorithm which queries
all items.

We improve on this naive algorithm by
re-using the information from previous queries.
Assume that we just checked if $v_S$ is marked
by querying all $x_i$, $i\in S$.  
If the next vertex $v_T$ is such that
$T$ contains only $m$ elements $i\notin S$, then
we only need to query $m$ elements $x_i$, $i\in T\setminus S$
instead of $r=N^{2/3}$ elements $x_i$, $i\in T$.

To formalize this, we use the following model.
At each moment, we are at one vertex of $G$
(superposition of vertices in quantum case).
In one time step, we can examine if the 
current vertex $v_S$ is marked and move to an adjacent vertex $v_T$.
Assume that there is an algorithm $A$ that finds a marked vertex 
with $M$ moves between vertices. 
Then, there is an algorithm
that solves element distinctness in $M+r$ steps, 
in a following way:
\begin{enumerate}
\item
We use $r$ queries to query all $x_i$, 
$i\in S$ for the starting vertex $v_S$.
\item
We then repeat the following two operations $M$ times:
\begin{enumerate}
\item
Check if the current vertex $v_S$ is marked.
This can be done without any queries because 
we already know all $x_i$, $i\in S$.
\item
We simulate the algorithm $A$ until the next move, find
the vertex $v_T$ to which it moves from $v_S$. 
We then move to $v_T$, by querying 
$x_i$, $i\in T\setminus S$.
After that, we know all $x_i, i\in T$.
We then set $S=T$.
\end{enumerate} 
\end{enumerate}

The total number of queries is at most $M+r$, consisting of 
$r$ queries for the first step and 1 query 
to simulate each move of $A$.

In the next sections, we will show how to search this
graph by quantum walk in $O(N^{2/3})$ steps
for element distinctness and $O(N^{k/(k+1)})$ steps
for $k$-distinctness.

\begin{Algorithm}
\begin{enumerate}
\item
\label{diffusion1}
Apply the transformation mapping $\ket{S}\ket{y}$ to
\[ \ket{S}\left(\left(-1+\frac{2}{N-r}\right) 
\ket{y}+\frac{2}{N-r}\sum_{y'\notin S, y'\neq y} \ket{y'}\right) .\]
on the $S$ and $y$ registers 
of the state in $\H$.
(This transformation is a variant of ``diffusion transformation''
in \cite{Grover}.)
\item
\label{step:add}
Map the state from $\H$ to $\H'$ by 
adding $y$ to $S$ and changing $x$ to a vector of
length $k+1$ by introducing 0 in the location
corresponding to $y$:
\item
Query for $x_y$ and insert it into location of $x$
corresponding to $y$.
\item
\label{diffusion2}
Apply the transformation mapping $\ket{S}\ket{y}$ to
\[  \ket{S}\left(
\left(-1+\frac{2}{r+1}\right)
\ket{y}+\frac{2}{r+1}\sum_{y'\in S, y'\neq y} \ket{y'}\right) .\]
on the $y$ register.
\item
\label{step:remove}
Erase the element of $x$ corresponding to new $y$ by
using it as the input to query for $x_y$.
\item
Map the state back to $\H$ by
removing the 0 component corresponding to $y$
from $x$ and removing $y$ from $S$.
\end{enumerate}
\caption{One step of quantum walk}
\label{alg:qw}
\end{Algorithm}

\begin{Algorithm}
\begin{enumerate}
\item
Generate the uniform superposition 
$\frac{1}{\sqrt{{N\choose r}(N-r)}} \sum_{|S|=r, y\notin S} \ket{S}\ket{y}$.
\item
Query all $x_i$ for $i\in S$. This transforms the state to 
\[ \frac{1}{\sqrt{{N\choose r}(N-r)}} \sum_{|S|=r, y\notin S} 
\ket{S}\ket{y} \bigotimes_{i\in S} \ket{x_i} .\]
\item 
\label{step}
$t_1=O((N/r)^{k/2})$ times repeat:
\begin{enumerate}
\item
\label{step1}
Apply the conditional phase flip 
(the transformation $\ket{S}\ket{y}\ket{x}\rightarrow -\ket{S}\ket{y}\ket{x}$) 
for $S$ such that $x_{i_1}=x_{i_2}=\ldots=x_{i_k}$ for  
$k$ distinct $i_1, \ldots, i_k \in S$.
\item
\label{step2}
Perform $t_2=O(\sqrt{r})$ steps of the quantum walk (algorithm \ref{alg:qw}).
\end{enumerate}
\item
Measure the final state. Check if $S$ contains a $k$-collision and 
answer ``there is a $k$-collision" or ``there is no $k$-collision",
according to the result. 
\end{enumerate}
\caption{Single-solution algorithm}
\label{MainAlg}
\end{Algorithm}

\subsection{The algorithm}

Let $x_1, \ldots, x_N\in [M]$. 
We consider two Hilbert spaces $\H$ and $\H'$.
$\H$ has dimension
${N\choose r}M^r (N-r)$ and
the basis states of $\H$ are
$\ket{S, x, y}$ with
$S\subseteq [N]$, $|S|=r$, 
$x\in[M]^r$, $y\in [N]\setminus S$.
$\H'$ has dimension ${N\choose r+1}M^{r+1} (r+1)$.
The basis states of $\H'$ are
$\ket{S, x, y}$ with
$S\subseteq [N]$, $|S|=r+1$, 
$x\in[M]^{r+1}$, $y\in S$.
Our algorithm thus uses
\[ O\left( {N\choose r}M^r (N-r)+{N\choose r+1}M^{r+1} (r+1) \right)=
O(r (\log N+\log M)) \]
qubits of memory.

In the states used by our algorithm,
$x$ will always be equal to 
$(x_{i_1}, \ldots, x_{i_r})$ 
where $i_1, \ldots, i_r$ are elements
of $S$ in increasing order.

We start by defining a quantum walk 
on $\H$ and $\H'$ (algorithm \ref{alg:qw}).
Each step of the quantum walk starts in a superposition
of states in $\H$. The first three steps map the
state from $\H$ to $\H'$ and the last three steps map it back
to $\H$. 

If there is at most one $k$-collision,
we apply Algorithm \ref{MainAlg}
($t_1$ and $t_2$ are $c_1\sqrt{r}$ and $c_2(\frac{N}{r})^{k/2}$
for constants $c_1$ and $c_2$ which can be calculated from the analysis
in section \ref{sec:MainAn}).
This algorithm alternates quantum walk with a transformation
that changes the phase if the current state contains a $k$-collision. 
We give a proof of correctness for Algorithm 
\ref{MainAlg} in section \ref{sec:MainAn}. 

 
If there can be more one $k$-collision,
element $k$-distinctness is solved by algorithm \ref{MultAlg}.
Algorithm \ref{MultAlg} is a classical
algorithm that randomly selects several subsets of $x_i$
and runs algorithm \ref{MainAlg} on each subset.
We give Algorithm \ref{MultAlg} and its analysis in
section \ref{sec:multiple}.


\section{Analysis of single $k$-collision algorithm}
\label{sec:MainAn}

\subsection{Overview}

The number of queries for algorithm \ref{MainAlg} is $r$ 
for creating the initial state
and $O((N/r)^{k/2} \sqrt{r})=O(\frac{N^{k/2}}{r^{(k-1)/2}})$ 
for the rest of the algorithm.
Thus, the overall number of queries is 
$O(\max (r, \frac{N^{k/2}}{r^{(k-1)/2}}))$.
The correctness of algorithm \ref{MainAlg} follows from

\begin{Theorem}
\label{thm:main}
Let the input $x_1$, $\ldots$, $x_N$ be such that $x_{i_1}=\ldots=x_{i_k}$ for exactly 
one set of $k$ distinct values $i_1, \ldots, i_k$. With a constant probability, 
measuring the final state of algorithm \ref{MainAlg} gives $S$ such that $i_1, \ldots, i_k\in S$.
\end{Theorem}

\proof
The main ideas are as follows.
We first
 show (Lemma \ref{lem:symmetry}) 
that algorithm's state always stays in a $2k+1$-dimensional
subspace of $\H$. After that (Lemma \ref{lem:onestep}), we find 
the eigenvalues for the unitary transformation
induced by one step of the quantum walk (algorithm \ref{alg:qw}), 
restricted to this subspace. We then look at algorithm \ref{MainAlg}
as a sequence of the form $(U_2 U_1)^{t_1}$ with $U_1$ being
a conditional phase flip and $U_2$ being a unitary transformation
whose eigenvalues have certain properties (in this case,
$U_2$ is $t_2$ steps of quantum walk).
We then prove a general result (Lemma \ref{lem:gengrover}) 
about such sequences, which implies that the algorithm finds 
the $k$-collision
 with a constant probability.  

Let $\ket{S, y}$ be a shortcut for the basis state 
$\ket{S}\otimes_{i\in S}\ket{x_i}\ket{y}$.
In our algorithm, the $\ket{x}$ register of a state $\ket{S, x, y}$ 
always contains the state $\otimes_{i\in S}\ket{x_i}$.
Therefore, the state of the algorithm is always
a linear combination of the basis states $\ket{S, y}$.

We classify the basis states $\ket{S, y}$ ($|S|=r$, $y\notin S$) into $2k+1$ types.
A state $\ket{S,y}$ is of type $(j, 0)$ if $|S\cap \{i_1, \ldots, i_k\}|=j$ and 
$y\notin\{i_1, \ldots, i_k\}$ and of type $(j, 1)$ 
if $|S\cap \{i_1, \ldots, i_k\}|=j$ and $y\in\{i_1, \ldots, i_k\}$. 
For $j\in\{0, \ldots, k-1\}$, there are both type $(j, 0)$ and 
type $(j, 1)$ states. For $j=k$, there are only $(k, 0)$ type states.
($(k, 1)$ type is impossible because, if, $|S\cap \{i_1, \ldots, i_k\}|=k$,
then $y\notin S$ implies $y\notin\{i_1, \ldots, i_k\}$.)

Let $\ket{\psi_{j ,l}}$ be the uniform superposition of basis states $\ket{S, y}$ of type $(j, l)$.
Let $\tilde{H}$ be the ($2k+1$)-dimensional space spanned by states $\ket{\psi_{j, l}}$.

For the space $\H'$, its basis states $\ket{S, y}$ ($|S|=r+1$, $y\in S$) 
can be similarly classified into $2k+1$ types.
We denote those types $(j, l)$ with $j=|S\cap\{i_1, \ldots, i_k\}|$, 
$l=1$ if $y\in\{i_1, \ldots, i_k\}$ and $l=0$ otherwise. 
(Notice that, since $y\in S$ for the space $\H'$, we have type $(k, 1)$ but no 
type $(0, 1)$.)
Let $\ket{\varphi_{j ,l}}$ be the uniform superposition of basis states $\ket{S, y}$ 
of type $(j, l)$ for space $\H'$.
Let $\tilde{H'}$ be the ($2k+1$)-dimensional space spanned by $\ket{\varphi_{j, l}}$.
Notice that the transformation $\ket{S, y}\rightarrow \ket{S\cup\{y\}, y}$
maps 
\[ \ket{\psi_{i, 0}}\rightarrow \ket{\varphi_{i, 0}},\mbox{~}
\ket{\psi_{i, 1}}\rightarrow \ket{\varphi_{i+1, 1}}.\]
We claim
\begin{Lemma}
\label{lem:symmetry}
In algorithm \ref{alg:qw}, steps 1-3 map $\tilde{\H}$ to $\tilde{\H'}$ and steps
4-6 map $\tilde{\H'}$ to $\tilde{\H}$.
\end{Lemma}

\proof
In section \ref{sec:proofs1}. 
\qed

Thus, algorithm \ref{alg:qw} maps $\tilde{\H}$ to itself.
Also, in algorithm \ref{MainAlg}, step \ref{step1} 
maps $\ket{\psi_{k, 0}}\rightarrow -\ket{\psi_{k, 0}}$ and 
leaves $\ket{\psi_{j, l}}$ for $j<k$ unchanged 
(because $\ket{\psi_{j, l}}$, $j<k$ are superpositions
of states $\ket{S, y}$ which are unchanged by step \ref{step2} and $\ket{\psi_{k, 0}}$ is
a superposition of states $\ket{S, y}$ which are mapped to 
$-\ket{S, y}$ by step \ref{step2}).
Thus, every step of algorithm \ref{MainAlg} maps $\tilde{\H}$ to itself.
Also, the starting state of algorithm \ref{MainAlg} can be expressed as
a combination of $\ket{\psi_{j, l}}$.
Therefore, it suffices to analyze algorithms \ref{alg:qw} and \ref{MainAlg} on
subspace $\tilde{\H}$.

In this subspace, we will be interested in two particular states.
Let $\ket{\psi_{start}}$ be the uniform superposition
of all $\ket{S, y}$, $|S|=r$, $y\notin S$.
Let $\ket{\psi_{good}}=\ket{\psi_{k, 0}}$ be the uniform superposition of
all $\ket{S, y}$ with $i_1, \ldots, i_k\in S$.
$\ket{\psi_{start}}$ is the algorithm's starting state.
$\ket{\psi_{good}}$ is the state we would like to obtain (because measuring
$\ket{\psi_{good}}$ gives a random set $S$ such that 
$\{i_1, \ldots, i_k\}\subseteq S$).

We start by analyzing a single step of quantum walk.

\begin{Lemma}
\label{lem:onestep}
Let $U$ be the unitary transformation induced on 
$\tilde{\H}$ by one step of the quantum walk
(algorithm \ref{alg:qw}). 
$U$ has $2k+1$ different eigenvalues in $\tilde{\H}$. One of them
is 1, with $\ket{\psi_{start}}$ being the eigenvector.
The other eigenvalues are $e^{\pm\theta_1 i}$, $\ldots$, $e^{\pm\theta_k i}$
with $\theta_j=(2\sqrt{j}+o(1))\frac{1}{\sqrt{r}}$.
\end{Lemma}

\proof
In section \ref{sec:proofs1}.
\qed

We set $t_2=\lceil\frac{\pi}{3\sqrt{k}} \sqrt{r}\rceil$. 
Since one step of quantum walk
fixes $\tilde{\H}$, $t_2$ steps fix $\tilde{\H}$ as well.
Moreover, $\ket{\psi_{start}}$ will still be an eigenvector 
with eigenvalue 1. The other $2k$ eigenvalues become
$e^{\pm i(\frac{2\pi\sqrt{j}}{3\sqrt{k}}+o(1))}$. 
Thus, every of those eigenvalues is 
$e^{i\theta}$ with $\theta\in[c, 2\pi-c]$,
for a constant $c$ independent of $N$ and $r$.

Let step $U_1$ be step \ref{step1} of algorithm \ref{MainAlg} and  
$U_2=U^{t_2}$ be step \ref{step2}. Then, the entire algorithm 
consists of applying $(U_2 U_1)^{t_1}$ to $\ket{\psi_{start}}$.
We will apply

\begin{Lemma}
\label{lem:gengrover}
Let $\H$ be a finite dimensional Hilbert space
and $\ket{\psi_1}$, $\ldots$, $\ket{\psi_m}$ be an orthonormal 
basis for $\H$.
Let $\ket{\psi_{good}}$, $\ket{\psi_{start}}$ be two states in $\H$ 
which are superpositions of $\ket{\psi_1}$, $\ldots$, $\ket{\psi_m}$
with real amplitudes and $\lbra \psi_{good} | \psi_{start}\rket=\alpha$.
Let $U_1$, $U_2$ be unitary transformations on $\H$ with
the following properties:
\begin{enumerate}
\item
$U_1$ is the transformation that flips the phase on 
$\ket{\psi_{good}}$ ($U_1\ket{\psi_{good}}=-\ket{\psi_{good}}$)
and leaves any state orthogonal to $\ket{\psi_{good}}$ unchanged.
\item
$U_2$ is a transformation which is described by a real-valued
$m\times m$ matrix in the basis $\ket{\psi_1}$, $\ldots$, $\ket{\psi_m}$.
Moreover, $U_2\ket{\psi_{start}}=\ket{\psi_{start}}$ 
and, if $\ket{\psi}$ is an 
eigenvector of $U_2$ perpendicular to $\ket{\psi_{start}}$, then
$U_2\ket{\psi}=e^{i\theta}\ket{\psi}$ for 
$\theta\in[\epsilon, 2\pi-\epsilon]$, $\theta\neq\pi$ (where $\epsilon$ is
a constant, $\epsilon>0$)\footnote{The requirement $\theta\neq\pi$
is made to simplify the proof of the lemma. The lemma remains true
if $\theta=\pi$ is allowed. At the end of section \ref{sec:proofs2},
we sketch how to modify the proof for this case.}
\end{enumerate} 
Then, there exists $t=O(\frac{1}{\alpha})$ such that 
$|\bra{\psi_{good}} (U_2 U_1)^{t} \ket{\psi_{start}}| =\Omega(1)$.
(The constant under $\Omega(1)$ is independent of $\alpha$ 
but can depend on $\epsilon$.)
\end{Lemma}

\proof
In section \ref{sec:proofs2}.
\qed

\comment{
{\bf Example.} 
A particular case of this lemma is Grover's algorithm.
Then, $\ket{\psi_{start}}=\frac{1}{\sqrt{N}} \sum_{i=1}^N \ket{i}$
and $\ket{\psi_{good}}$ is the basis state corresponding to the marked item
(or uniform superposition of all such basis states if there is
more than one marked item). 
$U_1$ is just the query transformation. $U_2$ is the "diffusion transform"
which maps $U_2\ket{\psi_{start}}=\ket{\psi_{start}}$ and $U_2\ket{\psi}=-\ket{\psi}$
for any $\ket{\psi}$ orthogonal to $\ket{\psi_{start}}$.
Thus, $e^{i\theta_i}=-1$ and $\theta_i=\pi$ for all $i$.
For one marked item,
$\alpha=\lbra s | w\rket=\frac{1}{\sqrt{N}}$.
Our Lemma implies that, after $O(\frac{1}{\alpha})=O(\sqrt{N})$
iterations, the amplitude of marked item is constant.

In our case, the Hilbert space is $\H_5$
and the basis is $\ket{\psi_{0, 0}}$, $\ldots$, $\ket{\psi_{2, 0}}$.
The transformations $U_1$ and $U_2$ are steps 
\ref{step1} and \ref{step2} of Algorithm \ref{MainAlg}.
$U_1$ changes phase on the state $\ket{\psi_{good}}=\ket{\psi_{2, 0}}$
and leaves any superposition of the other 4 basis states
unchanged. $U_2$ keeps the state $\ket{\psi_{start}}$ unchanged
and, for any other eigenvector, the eigenvalue is 
$e^{i\theta}$ with $\theta\geq \frac{2\pi}{3}-o(1)$.
Moreover, as seen from the proof of Lemma \ref{lem:onestep},
$U_2$ is described by a real matrix in the basis 
$\ket{\psi_{0, 0}}$, $\ldots$, $\ket{\psi_{2, 0}}$.
}

By Lemma \ref{lem:gengrover}, we can set $t_1=O(\frac{1}{\alpha})$ 
so that the inner product of $(U_2 U_1)^{t_1}\ket{\psi_{start}}$ and
$\ket{\psi_{good}}$ is a constant.
Since $\ket{\psi_{good}}$ is a superposition
of $\ket{S, y}$ over $S$ satisfying $\{i_1, \ldots, i_k\}\subseteq S$, 
measuring $(U_2 U_1)^{t_1}\ket{\psi_{start}}$
gives a set $S$ satisfying $\{i_1, \ldots, i_k\}\subseteq S$ with a constant
probability.

It remains to calculate $\alpha$. Let $\alpha'$ be the fraction
of $S$ satisfying $\{i_1, \ldots, i_k\}\subseteq S$.
Since $\ket{\psi_{start}}$ is the uniform superposition of
all $\ket{S, y}$ and $\ket{\psi_{good}}$ is the 
uniform superposition of $\ket{S, y}$ with $\{i_1, \ldots, i_k\}\subseteq S$
we have $\alpha=\sqrt{\alpha'}$. 
\[ \alpha'=Pr[\{i_1, \ldots, i_k\} \subseteq S]=  
\frac{{N-k \choose r-k}}{{N \choose r}} =
\frac{r}{N}\prod_{j=1}^{k-1} \frac{r-j}{N-j} =
(1-o(1))\frac{r^k}{N^k}. \] 
Therefore, $\alpha=\Omega(\frac{r^{k/2}}{N^{k/2}})$ and $t_1=O((N/r)^{k/2})$.
\qed

Lemma \ref{lem:gengrover} might also be interesting by itself.
It generalizes one of analyses of Grover's algorithm \cite{Aharonov}.
Informally, the lemma says that, in Grover-like sequence of transformations $(U_2 U_1)^t$,
we can significantly relax the constraints on $U_2$ and the algorithm
will still give similar result.
It is quite likely that such situations might appear
in analysis of other algorithms.

For the quantum walk for element $k$-distinctness,
Childs and Eisenberg \cite{CE} have improved the analysis 
of lemma \ref{lem:gengrover}, by showing that 
$\bra{\psi_{good}} (U_2 U_1)^{t} \ket{\psi_{start}}$
(and, hence, algorithm's success probability) is $1-o(1)$. 
Their result, however, does not apply to arbitrary transformations 
$U_1$ and $U_2$ satisfying conditions of lemma \ref{lem:gengrover}.

\subsection{Proofs of Lemmas \ref{lem:symmetry} and \ref{lem:onestep}}
\label{sec:proofs1}

\proof [of Lemma \ref{lem:symmetry}]
To show that $\tilde{\H}$ is mapped to $\tilde{\H'}$, it suffices
to show that each of basis vectors $\ket{\psi_{j, l}}$ is mapped
to a vector in $\tilde{\H'}$. Consider vectors $\ket{\psi_{j, 0}}$ 
and $\ket{\psi_{j, 1}}$ for $j\in \{0, 1, \ldots, k-1\}$.
Fix $S$, $|S\cap\{i_1, \ldots, i_k\}|=j$.
We divide $[N]\setminus S$ into two sets $S_0$ and $S_1$. Let
\[ S_0=\{ y: y\in[N]\setminus S, y\notin\{i_1, \ldots, i_k\} \} ,\]
\[ S_1=\{ y: y\in[N]\setminus S, y\in\{i_1, \ldots, i_k\} \} .\]
Since $|S \cap \{i_1, \ldots, i_k\}|=j$,
$S_1$ contains $s_1=k-j$ elements. 
Since $S_0\cup S_1=[N]\setminus S$ contains $N-r$ elements, $S_0$ contains
$s_0=N-r-k+j$ elements.
Define $\ket{\psi_{S, 0}}=\frac{1}{\sqrt{N-r-k+j}}\sum_{y\in S_0}\ket{S,y}$ and
$\ket{\psi_{S, 1}}=\frac{1}{\sqrt{k-j}}\sum_{y\in S_1}\ket{S,y}$.
Then, we have 
\begin{equation}
\label{eq-2110}
\ket{\psi_{j, 0}}=\frac{1}{\sqrt{{k\choose j}{N-k \choose r-j}}} 
\mathop{\sum_{S:|S|=r}}_{|S\cap\{i_1, \ldots, i_k\}|=j} 
\ket{\psi_{S, 0}}
\end{equation}
and, similarly for $\ket{\psi_{j, 1}}$ and $\ket{\psi_{S, 1}}$.

Consider the step 1 of algorithm \ref{alg:qw}, applied to
the state $\ket{\psi_{S, 0}}$. Let $\ket{\psi'_{S, 0}}$ be the resulting state.
Since the $\ket{S}$ register is
unchanged, $\ket{\psi'_{S, 0}}$ is some superposition of states $\ket{S, y}$.
Moreover, both the state $\ket{\psi_{S, 0}}$ and the transformation
applied to this state in step 1 are invariant under permutation of 
states $\ket{S, y}$, $y\in S_0$ or states $\ket{S, y}$, $y\in S_1$.
Therefore, the resulting state must be invariant under such permutations
as well. This means that every $\ket{S, y}$, $y\in S_0$ and every $\ket{S, y}$, $y\in S_1$
has the same amplitude in $\ket{\psi'_{S, 0}}$.
This is equivalent to $\ket{\psi'_{S, 0}}=a\ket{\psi_{S, 0}}+b\ket{\psi_{S, 1}}$
for some $a$, $b$. 
Because of equation (\ref{eq-2110}), this means that
step 1 maps $\ket{\psi_{j, 0}}$ to $a\ket{\psi_{j, 0}}+b\ket{\psi_{j, 1}}$.
Steps 2 and 3 then map $\ket{\psi_{j, 0}}$ to $\ket{\varphi_{j, 0}}$ and
$\ket{\psi_{j, 1}}$ to $\ket{\varphi_{j+1, 1}}$.
Thus, $\ket{\psi_{j, 0}}$ is mapped to a superposition of two basis
states of $\tilde{\H'}$: $\ket{\varphi_{j, 0}}$ and $\ket{\varphi_{j+1, 1}}$.
Similarly, $\ket{\psi_{j, 1}}$ is mapped to a (different)
superposition of those two states.

For $j=k$, we only have one state $\ket{\psi_{k, 0}}$. A similar
argument shows that this state is unchanged by step 1 and then mapped
to $\ket{\varphi_{k, 0}}$ which belongs to $\tilde{\H'}$.

Thus, steps 1-3 map $\tilde{\H}$ to $\tilde{\H'}$.
The proof that steps 4-6 map $\tilde{\H'}$ to $\tilde{\H}$
is similar.
\qed

\proof [of Lemma \ref{lem:onestep}]
We fix a  basis for $\tilde{\H}$ consisting of 
$\ket{\psi_{j,0}}$, $\ket{\psi_{j,1} }$, $j\in\{0, \ldots, k-1\}$
and $\ket{\psi_{k, 0}}$ and a basis for $\tilde{\H'}$ consisting of
$\ket{\varphi_{0, 0}}$ and $\ket{\varphi_{j, 1}}$, 
$\ket{\varphi_{j, 0}}$, $j\in\{1, \ldots, k\}$.
Let $D_{\epsilon}$ be the matrix
\[
D_{\epsilon}=\left(
\begin{array}{cc}
1-2\epsilon & 2\sqrt{\epsilon-\epsilon^2} \\
2\sqrt{\epsilon-\epsilon^2} & -1+2\epsilon
\end{array}
\right) .\]

\begin{Claim}
Let $U_1$ be the unitary transformation mapping $\tilde{\H}$ to $\tilde{\H'}$
induced by steps 1-3 of quantum walk.
Then, $U_1$ is described by a block diagonal matrix
\[ U_1=\left( 
\begin{array}{ccccc}
D_{\frac{k}{N-r}} & 0 & \ldots & 0 & 0 \\
0 & D_{\frac{k-1}{N-r}} & \ldots & 0 & 0 \\
\vdots & \vdots & \ddots & \vdots & \vdots \\
0 & 0 & \ldots & D_{\frac{1}{N-r}} & 0 \\
0 & 0 & \ldots & 0 & 1 
\end{array} \right),\]
where the columns are 
in the basis $\ket{\psi_{0, 0}}$, $\ket{\psi_{0, 1}}$,
$\ket{\psi_{1, 0}}$, $\ket{\psi_{1, 1}}$, $\ldots$, $\ket{\psi_{k, 0}}$
and the rows are in the basis $\ket{\varphi_{0, 0}}$, $\ket{\varphi_{1, 1}}$,
$\ket{\varphi_{1, 0}}$, $\ket{\varphi_{2, 1}}$, $\ldots$, 
$\ket{\varphi_{k, 1}}$, $\ket{\varphi_{k, 0}}$.
\end{Claim}

\proof 
Let $\H_j$ be the 2-dimensional subspace of
$\tilde{\H}$ spanned by $\ket{\psi_{j, 0}}$ and $\ket{\psi_{j, 1}}$.
Let $\H'_j$ be the 2-dimensional subspace of
$\tilde{\H'}$ spanned by $\ket{\varphi_{j, 0}}$ and $\ket{\varphi_{j+1, 1}}$.

From the proof of Lemma \ref{lem:symmetry}, we know that
the subspace $\H_j$ is mapped to the subspace $\H'_j$.
Thus, we have a block diagonal matrices with $2\times 2$ blocks
mapping $\H_j$ to $\H'_j$ 
and $1\times 1$ identity matrix mapping $\ket{\psi_{k, 0}}$ to
$\ket{\varphi_{k, 0}}$.
It remains to show that the transformation 
from $\H_j$ to $\H'_j$ is $D_{\frac{k-j}{N-r}}$.
Let $S$ be such that $|S\cap\{i_1, \ldots, i_k\}|=j$.
Let $S_0, S_1$, $\ket{\psi_{S, 0}}$, $\ket{\psi_{S, 1}}$ be
as in the proof of lemma \ref{lem:symmetry}.
\comment{
We divide $y\in[N]-S$ into two sets $S_0$ and $S_1$. Let
\[ S_0=\{ y: y\in[N]-S, y\notin\{i_1, \ldots, i_k\} \} ,\]
\[ S_1=\{ y: y\in[N]-S, y\in\{i_1, \ldots, i_k\} \} .\]
Since $|S \cap \{i_1, \ldots, i_k\}|=j$,
$S_1$ contains $s_1=k-j$ elements. 
Since $S_0\cup S_1=[N]-S$ contains $N-r$ elements, $S_0$ contains
$s_0=N-r-k+j$ elements.
Define $\ket{\psi_{S, 0}}=\frac{1}{\sqrt{k-j}}\sum_{y\in S_0}\ket{S,y}$ and
$\ket{\psi_{S, 1}}=\frac{1}{\sqrt{N-r-k+j}}\sum_{y\in S_1}\ket{S,y}$.
Then, we have 
\begin{equation}
\label{eq-2110a}
\ket{\psi_{j, 0}}=\frac{1}{\sqrt{{r \choose j}{N-r \choose k-j}}}
\mathop{\sum_{S:|S|=r,}}_{|S\cap\{i_1, \ldots, i_k\}|=j}\ket{\psi_{S, 0}}
\end{equation}
and, similarly for $\ket{\psi_{j, 1}}$ and $\ket{\psi_{S, 1}}$.
Fix a particular $S$ for which $|\{ i_1, \ldots, i_k\}\cap S|=j$.}
Then, step 1 of algorithm \ref{alg:qw} maps 
$\ket{\psi_{S, 0}}$ to
\[ \frac{1}{\sqrt{s_0}} \sum_{y\in S_0} 
\left( \left(-1+\frac{2}{N-r}\right) \ket{S, y} + \sum_{y'\neq y, y'\notin S} 
\frac{2}{N-r} \ket{S, y'} \right)  \]
\[= \frac{1}{\sqrt{s_0}} \left(-1+\frac{2}{N-r}+(s_0-1)\frac{2}{N-r} \right)
\sum_{y\in S_0} \ket{S, y} + s_0\frac{1}{\sqrt{s_0}} \frac{2}{N-r} 
\sum_{y\in S_1} \ket{S, y}  \]
\[ =\left(-1+\frac{2s_0}{N-r}\right) \ket{\psi_{S, 0}} +
\frac{2\sqrt{s_0 s_1}}{N-r} \ket{\psi_{S, 1}} .\]
By a similar calculation, 
$\ket{\psi_{S, 1}}$ is mapped to 
\[ \left(-1+\frac{2s_1}{N-r}\right) 
\ket{\psi_{S, 1}} + \frac{2\sqrt{s_0 s_1}}{N-r} \ket{\psi_{S, 0}} =
\left(1-\frac{2s_0}{N-r}\right) 
\ket{\psi_{S, 1}} + \frac{2\sqrt{s_0 s_1}}{N-r} \ket{\psi_{S, 0}} .\]
By substituting $s_0=N-r-k+j$ and $s_1=k-j$, 
we see that step 1 produces the transformation $D_{\frac{k-j}{N-r}}$ on
$\ket{\psi_{S, 0}}$ and $\ket{\psi_{S, 1}}$.
Since $\ket{\psi_{j, 0}}$ and $\ket{\psi_{j, 1}}$ 
are uniform superpositions of $\ket{\psi_{S, 0}}$
and $\ket{\psi_{S, 1}}$ over all $S$, step 1 also produces
the same transformation $D_{\frac{k-j}{N-r}}$
on $\ket{\psi_{j, 0}}$ and $\ket{\psi_{j, 1}}$.
Steps 2 and 3 just map $\ket{\psi_{j, 0}}$ to $\ket{\varphi_{j, 0}}$
and $\ket{\psi_{j, 1}}$ to $\ket{\varphi_{j+1, 1}}$.
\qed

Similarly, steps 4-6 give the transformation $U_2$ described by block-diagonal
matrix
\[ U_2=\left( 
\begin{array}{ccccc}
1 & 0 & 0 & \ldots & 0 \\
0 & D'_{\frac{1}{r+1}} & 0 & \ldots & 0 \\
0 & 0 & D'_{\frac{2}{r+1}} & \ldots & 0 \\
\vdots & \vdots & \vdots & \ddots & \vdots \\
0 & 0 & 0 & \ldots & D'_{\frac{k}{r+1}} 
\end{array} \right) .\]
from $\tilde{\H'}$ to $\tilde{\H}$.
Here, $D'_{\epsilon}$ denotes the matrix
\[
D'_{\epsilon}=\left(
\begin{array}{cc}
-1+2\epsilon & 2\sqrt{\epsilon-\epsilon^2} \\
2\sqrt{\epsilon-\epsilon^2} & 1-2\epsilon
\end{array}
\right) .\]

A step of quantum walk is
$U=U_2 U_1$. Let $V$ be the diagonal matrix with even entries on
the diagonal being -1 and odd entries being 1.
Since $V^2=I$, we have $U=U_2 V^2 U_1 = U'_2 U'_1$
for $U'_2=U_2 V$ and $U'_1=V U_1$.
Let 
\[ 
E_{\epsilon}=\left(
\begin{array}{cc}
1-2\epsilon & 2\sqrt{\epsilon-\epsilon^2} \\
-2\sqrt{\epsilon-\epsilon^2} & 1-2\epsilon
\end{array}
\right) .\]
\comment{, \]
\[ E'_{\epsilon}=\left(
\begin{array}{cc}
1-2\epsilon & 2\sqrt{\epsilon-\epsilon^2} \\
-2\sqrt{\epsilon-\epsilon^2} & 1-2\epsilon
\end{array}
\right) ,
\]}

Then, $U'_1$ and $U'_2$ are equal to $U_1$ and $U_2$,
with every $D_{\epsilon}$ or $D'_{\epsilon}$ replaced
by corresponding $E_{\epsilon}$.
7We will first diagonalize $U'_1$ and $U'_2$ separately
and then argue that eigenvalues of $U'_2 U'_1$ are almost
the same as eigenvalues of $U'_2$.

Since $U'_2$ is block diagonal, it suffices to diagonalize
each block. $1\times 1$ identity block has eigenvalue 1.
For a matrix $E_{\epsilon}$, its characteristic 
polynomial is $\lambda^2-(2-4\epsilon)\lambda+1=0$
and its roots are $1-2 \epsilon \pm 2\sqrt{\epsilon-\epsilon^2} i$.
For $\epsilon=o(1)$, this is equal to
$e^{\pm (2+o(1)) i \sqrt{\epsilon}}$.
Thus, the eigenvalues of $U'_2$ are 
1, and $e^{\pm (2+o(1))\frac{\sqrt{j}}{\sqrt{r+1}} i }$ for 
$j\in\{1, 2, \ldots, k\}$.
Similarly, the eigenvalues of $U'_1$ are
1, and $e^{\pm (2+o(1))\frac{\sqrt{j}}{\sqrt{N-r}} i }$ for 
$j\in\{1, 2, \ldots, k\}$.

To complete the proof, we use the following bound on the eigenvalues
of the product of two matrices which follows 
from Hoffman-Wielandt theorem in matrix analysis \cite{Algebra}. 

\begin{Theorem}
\label{thm:g1}
Let $A$ and $B$ be unitary matrices.
Assume that $A$ has eigenvalues $1+\delta_1$, $\ldots$,
$1+\delta_m$, $B$ has eigenvalues
$\mu_1$, $\ldots$, $\mu_m$ and $AB$ has eigenvalues
$\mu'_1$, $\ldots$, $\mu'_m$.
Then, 
\[ |\mu_j-\mu'_j|\leq \sum_{i=1}^m |\delta_i| \]
for all $j\in[m]$.
\end{Theorem}

\proof
In section \ref{sec:app1}.
\qed

Let $A=U'_1$ and $B=U'_2$. 
Since $|e^{\epsilon i}-1|\leq |\epsilon|$, 
each of $|\delta_i|$ is of order $O(\frac{1}{\sqrt{N-r}})$.
Therefore, their sum is of order $O(\frac{1}{\sqrt{N-r}})$ as well.
Thus, for each eigenvalue
of $U'_2$, there is a corresponding eigenvalue 
of $U'_2 U'_1$ that differs by at most
by $O(\frac{1}{\sqrt{N-r}})$.
The lemma now follows from $\frac{1}{\sqrt{N-r}}=o(\frac{1}{\sqrt{r+1}})$.
\qed

\subsection{Proof of Lemma \ref{lem:gengrover}}
\label{sec:proofs2}

We assume that 
$|\alpha|<c \epsilon^2$ for some sufficiently small positive constant $c$. 
Otherwise, we can just take $t=0$ and get
$|\bra{\psi_{good}}(U_2 U_1)^t |\psi_{start}\rket|  =
|\bra{\psi_{good}}\psi_{start}\rket|=|\alpha| \geq c\epsilon^2$.

Consider the eigenvalues of $U_2$. Since $U_2$ is described
by a real $m\times m$ matrix (in the basis $\ket{\psi_1}$, $\ldots$,
$\ket{\psi_m}$), its characteristic polynomial has real coefficients.
Therefore, the eigenvalues are 1, -1, $e^{\pm i\theta_1}$, $\ldots$,
$e^{\pm i\theta_l}$. From conditions of the lemma, we know
that the eigenvalue of $e^{i\pi}=-1$ never occurs.

Let $\ket{w_{j,+}}$, $\ket{w_{j,-}}$ be the eigenvectors of $U_2$
with eigenvalues $e^{i\theta_j}$, $e^{-i\theta_j}$.
Let $\ket{w_{j,+}}=\sum_{j'=1}^l c_{j, j'} \ket{\psi_{j'}}$.
Then, we can assume that 
$\ket{w_{j,-}}=\sum_{j'=1}^l c^{*}_{j, j'} \ket{\psi_{j'}}$.
(Since $U_2$ is a real matrix, taking $U_2\ket{w_{j,+}}=e^{i\theta_j}\ket{w_{j,+}}$
and replacing every number with its complex conjugate gives
$U_2\ket{w}=e^{-i\theta_j} \ket{w}$ for 
$\ket{w}=\sum_{j=1}^l c^{*}_{j, j'} \ket{\psi_{j'}}$.)

We write $\ket{\psi_{good}}$ in a basis consisting of eigenvectors of $U_2$:
\begin{equation}
\label{eq:2510} 
\ket{\psi_{good}}=\alpha \ket{\psi_{start}} + 
\sum_{j=1}^l ( a_{j,+}\ket{w_{j,+}} + 
a_{j,-}\ket{w_{j,-}} ) .
\end{equation}
W. l. o. g., assume that $\alpha$ is a positive real. 
(Otherwise, multiply $\ket{\psi_{start}}$
by an appropriate factor to make $\alpha$ a positive real.)

We can also assume that $a_{j, +}=a_{j, -}=a_j$, with
$a_j$ being a positive real number. (To see that, let
$\ket{\psi_{good}}=\sum_{j'=1}^l b_{j'} \ket{\psi_{j'}}$. 
Then, $b_{j'}$ are real 
(by the assumptions of Lemma \ref{lem:gengrover}).
We have $\lbra  w_{j,+} | \psi_{good} \rket 
=a_{j,+}=\sum_{j'=1}^l b_{j'} c^*_{j, j'}$
and $\lbra w_{j,-} |\psi_{good} \rket = a_{j, -}
=\sum_{j'=1}^l b_{j'} (c^*_{j, j'})^* 
= (\sum_{j'=1}^l b_{j'} c^*_{j, j'})^* = a_{j,+}^*$.
Multiplying $\ket{w_{j,+}}$ by $\frac{a_{j, +}^*}{|a_{j, +}|}$
and $\ket{w_{j,-}}$ by $\frac{a_{j, +}}{|a_{j, +}|}$ 
makes both $a_{j, +}$ and $a_{j, -}$ equal to
$\frac{a_{j, +}a_{j, +}^*}{|a_{j, +}|}=|a_{j, +}|$ 
which is a positive real.)

Consider the vector 
\begin{equation}
\label{eq:vbeta}
\ket{v_{\beta}}=\alpha \left( 1+i\ctg\frac{\beta}{2} \right) \ket{\psi_{start}}+ 
 \sum_{j=1}^l a_{j} \left( 1+i\ctg\frac{-\theta_j+\beta}{2} \right) \ket{w_{j,+}} + 
\sum_{j=1}^l a_{j} \left( 1+i\ctg\frac{\theta_j+\beta}{2} \right)\ket{w_{j,-}} .
\end{equation} 
We will prove that, for some $\beta=\Omega(\alpha)$, $\ket{v_{\beta}}$ and 
$\ket{v_{-\beta}}$ are eigenvectors of $U_2U_1$, with eigenvalues $e^{\pm i\beta}$.
After that, we show that the starting state $\ket{\psi_{start}}$ 
is close to the state $\frac{1}{\sqrt{2}}\ket{v_{\beta}}+
\frac{1}{\sqrt{2}}\ket{v_{-\beta}}$. Therefore, 
repeating $U_2 U_1$ $\frac{\pi}{2\beta}$ times transforms 
$\ket{\psi_{start}}$ to a state close to $\frac{i}{\sqrt{2}}\ket{v_{\beta}}+
\frac{-i}{\sqrt{2}}\ket{v_{-\beta}}$ which is equivalent
to $\frac{1}{\sqrt{2}}\ket{v_{\beta}}-\frac{1}{\sqrt{2}}\ket{v_{-\beta}}$.
We then complete the proof by showing that this state has a constant
inner product with $\ket{\psi_{good}}$.

We first state some bounds on trigonometric functions that
will be used throughout the proof.

\begin{Claim}
\label{claim:trig}
\begin{enumerate}
\item
\label{trig1}
$\frac{2x}{\pi}\leq \sin x\leq x$ for all $x\in[0, \frac{\pi}{2}]$;
\item
\label{trig2}
$\frac{\pi}{4x} \leq \cot x \leq \frac{1}{x}$ for all $x\in[0, \frac{\pi}{4}]$.
\end{enumerate}
\end{Claim}

We now start the proof by establishing a sufficient
condition for $\ket{v_{\beta}}$
and $\ket{v_{-\beta}}$ to be eigenvectors.
We have $\ket{v_{\beta}}=\ket{\psi_{good}}+i\ket{v'_{\beta}}$ where
\begin{equation}
\label{eq:vbeta1} 
\ket{v'_{\beta}}=\alpha \ctg\frac{\beta}{2} \ket{\psi_{start}}+ 
 \sum_{j=1}^l a_{j} \ctg\frac{-\theta_j+\beta}{2} \ket{w_{j,+}} + 
 \sum_{j=1}^l a_{j} \ctg\frac{\theta_j+\beta}{2} \ket{w_{j,-}} .
\end{equation}

\begin{Claim}
\label{claim:beta1}
If $\ket{v'_{\beta}}$ is orthogonal to $\ket{\psi_{good}}$, then
$\ket{v_{\beta}}$ is an eigenvector of $U_2 U_1$ with an 
eigenvalue of $e^{i\beta}$
and $\ket{v_{-\beta}}$ is an eigenvector of $U_2 U_1$ with an 
eigenvalue of $e^{-i\beta}$.
\end{Claim}

\proof
Since $\ket{v'_{\beta}}$ is orthogonal to $\ket{\psi_{good}}$, 
we have $U_1\ket{v'_{\beta}}=\ket{v'_{\beta}}$ and
$U_1\ket{v_{\beta}}=-\ket{\psi_{good}}+i\ket{v'_{\beta}}$.
Therefore, 
\[ U_2 U_1\ket{v_{\beta}} = \alpha 
\left( -1+i\ctg\frac{\beta}{2} \right) \ket{\psi_{start}}+ 
 \sum_{j=1}^l a_{j} e^{i\theta_j} 
 \left( -1+i\ctg\frac{-\theta_j+\beta}{2} \right)\ket{w_{j,+}} +\]
\[ \sum_{j=1}^l a_{j} e^{-i\theta_j} 
 \left( -1+i\ctg\frac{\theta_j+\beta}{2} \right)\ket{w_{j,-}} .\]
Furthermore,
\[ 1+i\ctg x= \frac{\sin x+ i \cos x}{\sin x}=\frac{e^{i(\frac{\pi}{2}-x)}
}{\sin x},\]
\[ -1+i\ctg x= \frac{-\sin x+ i \cos x}{\sin x}=\frac{e^{i(\frac{\pi}{2}+x)}
}{\sin x},\]
Therefore,
\[ \left( -1+i\ctg\frac{\beta}{2} \right) 
= e^{i\beta} \left( 1+i\ctg\frac{\beta}{2} \right) ,\]
\[ e^{i\theta_j}\left( -1+i\ctg\frac{-\theta_j+\beta}{2} \right)=
\frac{e^{i(\frac{\pi}{2}+\frac{\theta_j}{2}+\frac{\beta}{2}) }}{\sin 
\frac{-\theta_j+\beta}{2}}=
 e^{i\beta} 
\left( 1+i\ctg\frac{-\theta_j+\beta}{2} \right) \]
and similarly for the coefficient of $\ket{w_{j, -}}$.
This means that $U_2 U_1\ket{v_{\beta}} =e^{i\beta} \ket{v_{\beta}}$.

For $\ket{v_{-\beta}}$, we write out the inner products 
$\lbra \psi_{good} \ket{v'_{\beta}}$ and $\lbra \psi_{good} \ket{v'_{-\beta}}$.
Then, we see that $\lbra \psi_{good} \ket{v'_{-\beta}}=-\lbra \psi_{good} \ket{v'_{\beta}}$.
Therefore, if $\ket{\psi_{good}}$ and $\ket{v'_{\beta}}$ are orthogonal,
so are $\ket{\psi_{good}}$ and $\ket{v'_{-\beta}}$.
By the argument above, this implies that $\ket{v_{-\beta}}$ is
an eigenvector of $U_2U_1$ with an eigenvalue $e^{-i\beta}$.
\qed

Next, we use this necessary condition to bound $\beta$ for
which $\ket{v_{\beta}}$ and $\ket{v_{-\beta}}$ are eigenvectors.

\begin{Claim}
\label{claim:beta}
There exists $\beta$ such that
$\ket{v'_{\beta}}$ is orthogonal to $\ket{\psi_{good}}$
and $\frac{\epsilon \alpha}{\sqrt{\pi}}\leq \beta \leq 2.6 \alpha$.
\end{Claim}

\proof
Let $f(\beta)=\lbra \psi_{good} \ket{v'_{\beta}}$.
We have 
\[ f(\beta)= \alpha^2 \ctg\frac{\beta}{2} + \sum_{j=1}^l |a_j|^2 
\left( \ctg \frac{-\theta_j+\beta}{2}+\ctg \frac{\theta_j+\beta}{2} \right) .\]
We bound $f(\beta)$ from below and above, 
for $\beta\in[0, \frac{\epsilon}{2}]$.
For the first term, we have $\frac{\pi}{2\beta} \leq \ctg\frac{\beta}{2} \leq
\frac{2}{\beta}$ (by claim \ref{claim:trig}).
For the second term, we have
\begin{equation}
\label{eq:secondterm} 
\ctg \frac{-\theta_j+\beta}{2}+\ctg \frac{\theta_j+\beta}{2} = 
 -\frac{\sin{\beta}}{
\sin \frac{\theta_j+\beta}{2} \sin \frac{\theta_j-\beta}{2} } 
.\end{equation}
For the numerator, we have $\frac{2 \beta}{\pi} \leq \sin \beta\leq \beta$,
because of Claim \ref{claim:trig}.
The denominator can be bounded from below as follows:
\[ \sin \frac{\theta_j+\beta}{2} \sin \frac{\theta_j-\beta}{2}
\geq \sin\frac{\epsilon}{2} \sin\frac{\epsilon}{4} \geq
\frac{\epsilon^2}{2\pi^2} ,\]
with the first inequality following
from $\theta_j\geq \epsilon$ and $\beta\leq \frac{\epsilon}{2}$
and the last inequality following from 
claim \ref{claim:trig}. 
This means 
\begin{equation}
\label{eq:fbound} 
\alpha^2 \frac{\pi}{2 \beta} - \frac{(1-\alpha^2)\pi^2}{\epsilon^2} \beta 
\leq f(\beta) \leq \alpha^2 \frac{2}{\beta}- \frac{1-\alpha^2}{\pi}\beta ,
\end{equation}
where we have used $\|\psi_{good}\|^2=|\alpha|^2 + 2 \sum_{j=1}^l |a_j|^2$
(by equation (\ref{eq:2510})) and $\|\psi_{good}\|=1$ to replace
$\sum_{j=1}^l |a_j|^2$ by $\frac{1-\alpha^2}{2}$.

The lower bound of equation (\ref{eq:fbound}) 
implies that $f(\beta)\geq 0$ for 
$\beta=\frac{\epsilon}{\sqrt{2\pi(1-\alpha^2)}} \alpha$.
The upper bound implies that $f(\beta)\leq 0$ for
$\beta=\frac{\sqrt{2\pi}}{\sqrt{1-\alpha^2}} \alpha$.
Since $f$ is continuous, it must be the case that $f(\beta)=0$
for some $\beta\in[\frac{\epsilon}{\sqrt{2\pi(1-\alpha^2)}} \alpha,
\frac{\sqrt{2\pi}}{\sqrt{1-\alpha^2}} \alpha]$.
The claim now follows from $0\leq \alpha\leq 0.1$.
\qed

Let
$\ket{u_1}=\frac{ \ket{v_{\beta}} }{\|v_{\beta}\|}$ and
$\ket{u_2}=\frac{ \ket{v_{-\beta}} }{\|v_{-\beta}\|}$.
We show that $\ket{\psi_{start}}$ is almost a linear combination
of $\ket{u_1}$ and $\ket{u_2}$.
Define $\ket{\psi_{end}}=\frac{\ket{v_{end}}}{\|v_{end}\|}$ where 
\begin{equation}
\label{eq:vend} 
\ket{v_{end}}=\sum_{j=1}^l a_j \left( 1+i\ctg \frac{-\theta_j}{2} \right) \ket{w_{j, +}} +
\sum_{j=1}^l a_j \left( 1+i\ctg \frac{\theta_j}{2} \right) \ket{w_{j, -}} .
\end{equation}

\begin{Claim}
\label{claim:beta2}
\[ \ket{u_1}=c_{start} i \ket{\psi_{start}}+ c_{end} \ket{\psi_{end}} + \ket{u'_1} ,\]
\[ \ket{u_2}=-c_{start} i \ket{\psi_{start}}+ c_{end} \ket{\psi_{end}} + \ket{u'_2} \]
where $c_{start}$, $c_{end}$ are positive real numbers and $u'_1$, $u'_2$ satisfy 
$\|u'_1\|\leq \frac{3\beta}{\epsilon}$ 
and $\|u'_2\|\leq \frac{3\beta}{\epsilon}$,
for $\beta$ from Claim \ref{claim:beta}.
\end{Claim}

\proof
By regrouping terms in equation (\ref{eq:vbeta}), we have
\begin{equation}
\label{eq:2310} 
\ket{v_{\beta}}= \alpha i\ctg \frac{\beta}{2} \ket{\psi_{start}} 
+ \ket{v_{end}} + \ket{v''_{\beta}} 
\end{equation}
where 
\[ \ket{v''_{\beta}}= \alpha \ket{\psi_{start}} +
\sum_{j=1}^l a_j i \left(\ctg \frac{-\theta_j+\beta}{2} - \ctg \frac{-\theta_j}{2} \right)\ket{w_{j, +}} \]
\[ + \sum_{j=1}^l a_j i \left(\ctg \frac{\theta_j+\beta}{2} - \ctg \frac{\theta_j}{2} \right)\ket{w_{j, -}} .\]
We claim that $\|v''_{\beta}\|\leq \frac{3\beta}{\epsilon} \|v_{\beta}\|$.
We prove this by showing that the absolute value of each of coefficients in 
$\ket{v''_{\beta}}$ is at most $\frac{3\beta}{\epsilon}$ times the absolute 
value of corresponding coefficient in $\ket{v_{\beta}}$.
The coefficient of $\ket{\psi_{start}}$ is $\alpha$ in $\ket{v''_{\beta}}$ and 
$\alpha (1+i\ctg\frac{\beta}{2})$ in $\ket{v_{\beta}}$.
We have 
\[ |\alpha (1+i\ctg\frac{\beta}{2})| \geq \alpha \ctg \frac{\beta}{2} 
\geq \alpha \frac{8}{\pi \beta} ,\]
which means that the absolute value of the 
coefficient of $\ket{\psi_{start}}$ in 
$\ket{v''_{\beta}}$ is at most $\frac{\pi\beta}{8}$ times
the absolute value of the coefficient in $\ket{v_{\beta}}$. 
For the coefficient of the $\ket{w_{j, +}}$, we have 
\[\ctg \frac{-\theta_j+\beta}{2} - \ctg \frac{-\theta_j}{2} =
%
\frac{\sin\frac{\beta}{2}}{\sin \frac{-\theta_j+\beta}{2} \sin \frac{-\theta_j}{2} } \]
If $\theta_j-\beta\geq\frac{\pi}{2}$,
then
\[ \left| \frac{\sin\frac{\beta}{2}}{\sin \frac{-\theta_j+\beta}{2} 
\sin \frac{-\theta_j}{2}} \right| \leq 
\frac{\frac{\beta}{2}}{\sin \frac{\pi}{4} 
\sin \frac{\pi}{4}} = \frac{\frac{\beta}{2}}{
\frac{1}{\sqrt{2}}\frac{1}{\sqrt{2}}} = \beta \leq 
\beta\left|1+ i\ctg \frac{-\theta_j+\beta}{2}\right| .\]
If $\theta_j-\beta \leq \frac{\pi}{2}$, then 
\[ \left| \frac{\sin\frac{\beta}{2}}{\sin \frac{-\theta_j+\beta}{2} 
\sin \frac{-\theta_j}{2}} \right| = \left| \frac{\sin\frac{\beta}{2}}{
\cos \frac{-\theta_j+\beta}{2} \sin\frac{-\theta_j}{2}} 
\ctg \frac{-\theta_j+\beta}{2} \right| \leq 
\frac{\frac{\beta}{2}}{\frac{1}{\sqrt{2}} \frac{\theta_j}{\pi}} 
\ctg  \left| \frac{-\theta_j+\beta}{2} \right|\leq 
3\frac{\beta}{\epsilon}  \left| \ctg \frac{-\theta_j+\beta}{2} \right|,\]
with the first inequality following from
$|\cos \frac{-\theta_j+\beta}{2}| \geq |\cos\frac{\pi}{4}| =\frac{1}{\sqrt{2}}$
and $|\sin x| =\sin |x| \geq \frac{2|x|}{\pi}$ (using Claim \ref{claim:trig}).
Therefore, the absolute value of coefficient of $\ket{w_{j, +}}$ in 
$\ket{v''_{\beta}}$ is at most $\frac{3\beta}{\epsilon}$ times the absolute value 
of the coefficient of $\ket{w_{j, +}}$ in $\ket{v_{\beta}}$
(which is $|a_j(1+i\ctg \frac{-\theta_j+\beta}{2})|$). 
Similarly, we can bound the absolute value of coefficient of $\ket{w_{j, -}}$.

By dividing equation (\ref{eq:2310}) by $\|v_{\beta}\|$, we get
\[ \ket{u_1}= c_{start} i \ket{\psi_{start}} +
c_{end} \ket{\psi_{end}}+ \ket{u'_1} \]
for $c_{start}=\frac{\alpha \ctg \frac{\beta}{2}}{\|v_{\beta}\|}$,
$c_{end}=\frac{\|v_{end}\|}{\|v_{\beta}\|}$ 
and $\ket{u'_1}=\frac{1}{\|v_{\beta}\|} \ket{v''_{\beta}}$.
Since $\|v''_{\beta}\|\leq\frac{3 \beta}{\epsilon}  \|v_{\beta}\|$, 
we have 
$\|u'_1\|\leq\frac{3 \beta}{\epsilon}$.
The proof for $u_2$ is similar.
\qed

Since $\ket{u_1}$ and $\ket{u_2}$ are eigenvectors of
$U_2U_1$ with different eigenvalues, they must be orthogonal.
Therefore, 
\[ \lbra u_1 \ket{u_2}=-c_{start}^2+c_{end}^2+
O(\frac{\beta}{\epsilon})=0, \]
where $O(\frac{\beta}{\epsilon})$ denotes a term that is at
most $const \frac{\beta}{\epsilon}$ in absolute value 
for some constant $const$ that does not depend on $\beta$ and $\epsilon$.
Also,
\[ \|u_1\|^2=c_{start}^2+c_{end}^2+O(\frac{\beta}{\epsilon}) = 1 .\]
These two equalities together with $c_{start}$ and $c_{end}$ being
positive reals imply that 
$c_{start}=\frac{1}{\sqrt{2}}+O(\beta/\epsilon)$ and 
$c_{end}=\frac{1}{\sqrt{2}}+O(\beta/\epsilon)$.
Therefore, 
\[ \ket{u_1}=\frac{1}{\sqrt{2}} i \ket{\psi_{start}}+ 
 \frac{1}{\sqrt{2}} \ket{\psi_{end}} + \ket{u''_1} ,\]
\[ \ket{u_2}=-\frac{1}{\sqrt{2}} i \ket{\psi_{start}}+ 
 \frac{1}{\sqrt{2}}\ket{\psi_{end}} + \ket{u''_2}, \]
with $\|u''_1\|=O(\beta/\epsilon)$ and $\|u''_2\|=O(\beta/\epsilon)$.
This means that
\[ \ket{\psi_{start}}= -\frac{i}{\sqrt{2}} \ket{u_1}+\frac{i}{\sqrt{2}} \ket{u_2}+ \ket{w'} ,\]
\[ \ket{\psi_{end}}= \frac{1}{\sqrt{2}} \ket{u_1}+\frac{1}{\sqrt{2}} \ket{u_2}+ \ket{w''} ,\]
where $w'$ and $w''$ are states with $\|w'\|=O(\beta/\epsilon)$ 
and $\|w''\|=O(\beta/\epsilon)$.
Let $t=\lfloor \frac{\pi}{2\beta} \rfloor$.
Then, $(U_2 U_1)^t \ket{u_1}$ is almost $i \ket{u_1}$ (plus a 
term of order $O(\beta)$) and $(U_2 U_1)^t \ket{u_2}$ 
is almost $-i \ket{u_2}$.
Therefore,
\[ (U_2 U_1)^t \ket{\psi_{start}}=\ket{\psi_{end}}+\ket{v'} \]
where $\|v'\|=O(\beta/\epsilon)$.
This means that 
\begin{equation}
\label{eq:prob}
|\lbra \psi_{good} | (U_2 U_1)^t \ket{\psi_{start}}|
\geq |\lbra \psi_{good} \ket{\psi_{end}}|- O(\frac{\beta}{\epsilon}). 
\end{equation}
Since $\beta\leq 2.6 \alpha$ and $\alpha=c \epsilon^2$,
we have $O(\beta/\epsilon)=O(\epsilon)$. By choosing $c$ to be 
sufficiently small, we can make the $O(\beta/\epsilon)$ term
to be less than $0.1\epsilon$.
Then, Lemma \ref{lem:gengrover} follows from 

\begin{Claim}
\label{claim:beta3}
\[ |\lbra \psi_{good} \ket{\psi_{end}}|\geq 
\min\left( \frac{1-\alpha^2}{2}, \frac{1-\alpha^2}{4}\epsilon\right). \]
\end{Claim}

\proof
Since $\ket{\psi_{end}}=\frac{ \ket{v_{end}} }{\|v_{end}\|}$, we have
$\lbra\psi_{good} \ket{\psi_{end}}=\frac{ \lbra 
\psi_{good}\ket{v_{end}} }{\|v_{end}\|}$.
By definition of $\ket{v_{end}}$ (equation (\ref{eq:vend})),
$\lbra \psi_{good}\ket{v_{end}}=2 \sum_{j=1}^l a^2_j$.
By equation (\ref{eq:2510}), $\|\psi_{good}\|^2=\alpha^2+2 \sum_{j=1}^l a^2_j$.
Since $\|\psi_{good}\|^2=1$, 
we have $\lbra \psi_{good}\ket{v_{end}}=1-\alpha^2$.
Therefore,
$\lbra \psi_{good}\ket{\psi_{end}}\geq \frac{1-\alpha^2}{\|v_{end}\|}$.

We have $\|v_{end}\|^2 = 2 \sum_{j=1}^l a_j^2 
(1+\ctg^2 \frac{\theta_j}{2})$.
Since $\theta_k\in[\epsilon, 2\pi-\epsilon]$, 
$\| v_{end}\|^2\leq 2 \sum_{j=1}^l a^2_j (1+\ctg^2\frac{\epsilon}{2}) 
\leq (1+\ctg^2 \frac{\epsilon}{2})$
and 
\[ \lbra \psi_{good} \ket{\psi_{end}}\geq 
\frac{1-\alpha^2}{\sqrt{1+\ctg^2 (\epsilon/2)}}
\geq \frac{1-\alpha^2}{2 \max(1, \ctg \frac{\epsilon}{2})}
\geq \min\left( \frac{1-\alpha^2}{2}, \frac{1-\alpha^2}{4}\epsilon\right) .\]
\qed

If $\alpha$ is set to be sufficiently small, 
$|\lbra \psi_{good} \ket{\psi_{end}}|$ is close to $0.5\epsilon$
and, together with equation (\ref{eq:prob}), this means 
that $|\lbra \psi_{good} | (U_2 U_1)^t \ket{\psi_{start}}|$
is of order $\Omega(\epsilon)$.
\qed

\noindent
{\bf Remark.}
If $U_2$ has eigenvectors with eigenvalue -1, 
the equation (\ref{eq:2510}) becomes
\[ 
\ket{\psi_{good}}=\alpha \ket{\psi_{start}} + \sum_{j=1}^l 
( a_{j,+}\ket{w_{j,+}} + 
a_{j,-}\ket{w_{j,-}} )+a_{l+1} \ket{w_{l+1}} ,
\]
with $\ket{w_{l+1}}$ being an eigenvector with eigenvalue -1.
We also add $a_{l+1} (1-i \tan \frac{\beta}{2})\ket{w_{l+1}}$,
$-a_{l+1}i \tan \frac{\beta}{2}\ket{w_{l+1}}$ and 
$a_{l+1}\ket{w_{l+1}}$ terms
to the right hand sides of equations (\ref{eq:vbeta}), (\ref{eq:vbeta1})
and (\ref{eq:2310}), respectively.
Claims \ref{claim:beta1}, \ref{claim:beta}, \ref{claim:beta2} 
and \ref{claim:beta3} remain true, but proofs of claims 
require some modifications to handle
the $\ket{w_{l+1}}$ term.

\subsection{Derivation of Theorem \ref{thm:g1}}
\label{sec:app1}

In this section, we derive Theorem \ref{thm:g1} (which was used in the proof
of Lemma \ref{lem:onestep}) from Hoffman-Wielandt inequality.

\begin{Definition}
For a matrix $C=(c_{ij})$, we define its $l_2$-norm as 
$\|C\|=\sqrt{\sum_{i,j} |c^2_{ij}|}$.
\end{Definition}

\begin{Theorem}
\label{thm:g0}
\cite[pp. 292]{Algebra}
If $U$ is unitary, then $\|UC\|=\|C\|$ for any $C$.
\end{Theorem}

\begin{Theorem}
\label{thm:g}
\cite[Theorem 6.3.5]{Algebra}
Let $C$ and $D$ be $m\times m$ matrices. Let $\mu_1$, $\ldots$, $\mu_m$ and
$\mu'_1, \ldots, \mu'_m$ be eigenvalues of
$C$ and $D$, respectively. Then,
\[ \sum_{i=1}^m (\mu_i-\mu'_i)^2 \leq \|C-D\|^2 .\]
\end{Theorem}

To derive theorem \ref{thm:g1} from theorem \ref{thm:g}, let $C=B$ and $D=AB$.
Then, $C-D=(I-A) B$.
Since $B$ is unitary, $\|C-D\|=\|I-A\|$ (Theorem \ref{thm:g0}).
Let $U$ be a unitary matrix that diagonalizes $A$.
Then, $U (I-A) U^{-1}=I-UAU^{-1}$ and 
$\|I-A\|=\|I-UAU^{-1}\|$.
Since $UAU^{-1}$ is a diagonal matrix with $1+\delta_i$ on the diagonal,
$I-UAU^{-1}$ is a diagonal matrix with $\delta_i$ on the diagonal and
$\|I-UAU^{-1}\|^2=\sum_{i=1}^m |\delta_i|^2$
By applying Theorem \ref{thm:g} to $I$ and $UAU^{-1}$, we get
\[ \sum_{i=1}^m (\mu_i-\mu'_i)^2 \leq  \sum_{i=1}^m |\delta_i|^2 .\]
In particular, for every $i$, we have 
$(\mu_i-\mu'_i)^2 \leq  (\sum_{i=1}^m |\delta_i|^2 )$
and 
\[ |\mu_i-\mu'_i|\leq\sqrt{\sum_{i=1}^m |\delta_i|^2}\leq \sum_{i=1}^m |\delta_i| .\]

\section{Analysis of multiple $k$-collision algorithm}
\label{sec:multiple}

To solve the general case of $k$-distinctness, we run
Algorithm \ref{MainAlg} several times, on subsets of
the input $x_i, i\in [N]$. 

The simplest approach is as follows.
We first run Algorithm \ref{MainAlg} on the entire
input $x_i, i\in [N]$. We then chose a sequence of subsets
$T_1\subseteq [N]$, $T_2\subseteq [N]$, $\ldots$ 
with $T_i$ being a random subset of size $|T_i|=(\frac{2k}{2k+1})^i N$,
and run Algorithm \ref{MainAlg} on $x_i, i\in T_1$, then on
$x_i, i\in T_2$ and so on. It can be shown that, 
if the input $x_i, i\in [N]$ contains a $k$-collision, then
with probability at least 1/2, there exists $j$ such that
$x_i, i\in  T_j$ contains exactly one $k$-collision.
This means that running algorithm \ref{MainAlg}
on $x_i, i\in  T_j$ finds the $k$-collision with
a constant probability.

The difficulty with this solution is choosing subsets $T_j$.
If we chose a subset of size $\frac{2k}{2k+1} N$ uniformly
at random, we need $\Omega(N)$ space to store the subset
and $\Omega(N)$ time to generate it. Thus, the straightforward
implementation of this solution is efficient in terms of 
query complexity but not in terms of time or space. 
Algorithm \ref{MultAlg} is a more complicated implementation
of the same approach that also achieves time-efficiency
and space-efficiency.

\begin{Algorithm}
\begin{enumerate}
\item
Let $T_1=[N]$. Let $j=1$.
\item
While $|T_j|>\max(r, \sqrt{N})$ repeat:
\begin{enumerate}
\item
Run Algorithm \ref{MainAlg} on $x_i$, $i\in T_j$, using memory size
$r_j=\frac{r |T_j|}{N}$. 
Measure the final state, obtaining a set $S$.
If there are $k$ equal elements $x_i$, $i\in S$, stop, 
answer ``there is a $k$-collision''.
\item
Let $q_j$ be an even power of a prime with 
$|T_j|\leq q_j\leq (1+\frac{1}{2k^2}) |T_j|$.
Select a random permutation $\pi_j$ on $[q_j]$ from an
$\frac{1}{N}$-approximately $2k\log N$-wise independent 
family of permutations
(Theorem \ref{thm:kwise2}).
\item
Let 
\[ T_{j+1}=\left\{\pi^{-1}_1 \pi^{-1}_2 \ldots \pi^{-1}_{j} (i), 
i\in \left[\left\lceil\frac{2k}{2k+1} q_j \right\rceil\right]\right\} .\]
\item
Let $j=j+1$;
\end{enumerate}
\item
If $|T_j|\leq r$, query all $x_i$, $i\in T_j$ classically. If $k$ equal elements are found, answer 
``there is a $k$-collision'',
otherwise, answer ``there is no $k$-collision''.
\item
If $|T_j|\leq \sqrt{N}$, run Grover search on the set of at most $N^{k/2}$ 
$k$-tuples $(i_1, \ldots, i_k)$ of pairwise distinct $i_1, \ldots, i_k\in T_j$, searching
for a tuple $(i_1, \ldots, i_k)$ such that $x_{i_1}=\ldots=x_{i_k}$.
If such a tuple is found, answer 
``there is a $k$-collision'',
otherwise, answer ``there is no $k$-collision''.
\end{enumerate}
\caption{Multiple-solution algorithm}
\label{MultAlg}
\end{Algorithm}

We claim
\begin{Theorem}
\label{thm:mult}
\begin{enumerate}
\item[(a)]
Algorithm \ref{MultAlg} uses $O(r+\frac{N^{k/2}}{r^{(k-1)/2}})$ queries.
\item[(b)]
Let $p$ be the success probability of algorithm \ref{MainAlg}, if there is
exactly one $k$-collision. For any $x_1, \ldots, x_N$ 
containing at least one $k$-collision, 
algorithm \ref{MultAlg} finds a $k$-collision with probability 
at least $(1-o(1))p/2$.
\end{enumerate}
\end{Theorem}

\proof

{\bf Part (a).}
The second to last step of algorithm \ref{MultAlg} use at most $r$ queries. 
The last step uses $O(N^{k/4})$ queries and is performed 
only if $\sqrt{N}\geq r$. In this case, 
$\frac{N^{k/2}}{r^{(k-1)/2}}\geq \frac{N^{k/2}}{N^{(k-1)/4}}\geq N^{k/4}$. 
Thus, the last two steps use $O(r+\frac{N^{k/2}}{r^{(k-1)/2}})$ queries
and it suffices to show that
algorithm \ref{MultAlg} uses $O(r+\frac{N^{k/2}}{r^{(k-1)/2}})$ queries 
in its second step (the while loop).

Let $T_j$ and $r_j$ be as in algorithm \ref{MultAlg}.
Then $|T_1|=N$ and $|T_{j+1}|\leq \frac{2k}{2k+1} (1+\frac{1}{2k^2}) |T_j|$.
The number of queries in the $j^{\rm th}$ iteration of the while loop is 
of the order
\[
\frac{|T_j|^{k/2}}{r_j^{(k-1)/2}}+ r_j =
\frac{|T_j|^{k/2}}{(|T_j|r/N)^{(k-1)/2}}+ \frac{|T_j|r}{N} =
\frac{N^{(k-1)/2}}{r^{(k-1)/2}} \sqrt{|T_j|} + \frac{|T_j|r}{N} .
\]
The total number of queries in the while loop is of the order
\[ \sum_j
\left( \frac{N^{(k-1)/2}}{r^{(k-1)/2}} \sqrt{|T_j|} + \frac{|T_j|r}{N} 
\right) \leq
 \sum_{j=0}^{\infty} 
\left( \left( \frac{2k}{2k+1} \frac{2k^2+1}{2k^2}\right)^{j/2}
\frac{N^{k/2}}{r^{(k-1)/2}} +
\left( \frac{2k}{2k+1} \frac{2k^2+1}{2k^2}\right)^{j} r \right) \]
\begin{equation}
\label{eq:multalg} 
= O\left(\frac{N^{k/2}}{r^{(k-1)/2}}+r\right) .
\end{equation}

\noindent
{\bf Part (b).}
If $x_1, \ldots, x_N$ contain exactly one $k$-collision, then running
algorithm \ref{MainAlg} on all of $x_1, \ldots, x_N$ finds the $k$-collision with
probability at least $p$.
If $x_1, \ldots, x_N$ contain more than 
one $k$-collision, we can have three cases:
\begin{enumerate}
\item
For some $j$, $T_j$ contains more than one $k$-collision but $T_{j+1}$ contains exactly one $k$-collision.
\item
For some $j$, $T_j$ contains more than one $k$-collision but $T_{j+1}$ contains no 
$k$-collisions.
\item
All $T_j$ contain more than one $k$-collision (till $|T_j|$ becomes smaller than 
$\max(r, \sqrt{N})$
and the loop is stopped).
\end{enumerate}
In the first case, performing algorithm \ref{MainAlg} on $x_j$, $j\in T_{i+1}$
finds the $k$-collision with probability at least $p$.
In the second case, we have no guarantees about the probability at all.
In the third case, the last step of algorithm \ref{MultAlg} finds
one of $k$-collisions with probability 1.

We will show that the probability of the second case is always less than the
probability of the first case plus an asymptotically small quantity. 
This implies that, with probability at least $1/2-o(1)$,
either first or third case occurs. Therefore, the probability of
algorithm \ref{MultAlg} finding a $k$-collision is at least $(1/2-o(1))p$.
To complete the proof, we show

\begin{Lemma}
\label{lem:sampling}
Let $T$ be a set containing a $k$-collision.
Let $None_j$ be the event that $x_i, i\in T_j$ contains no $k$-collision and
$Unique_j$ be the event that $x_i, i\in T_j$ contains a unique $k$-collision.
Then, 
\begin{equation}
\label{eq:sampling}
Pr[Unique_{j+1}|T_j=T ] > Pr[None_{j+1}|T_j=T]-
o\left(\frac{1}{N^{1/4}}\right) 
\end{equation}
where $Pr[Unique_{j+1}|T_j=T ]$ and $Pr[None_{j+1}|T_j=T]$ denote
the conditional probabilities of $Unique_{j+1}$ and $None_{j+1}$,
if $T_j=T$.
\end{Lemma}

The probability of the first case is just the sum of probabilities 
\[ Pr[Unique_{j+1} \wedge T_j=T]=Pr[T_j=T] Pr[Unique_{j+1} | T_j=T] \]
over all $j$ and $T$ such that $|T|>\max(r, \sqrt{N})$ and 
$T$ contains more than one $k$-collision.
The probability of the second case is a similar sum of probabilities
\[ Pr[None_{j+1} \wedge T_j=T]=Pr[T_j=T] Pr[None_{j+1} | T_j=T] .\]
Therefore, $Pr[Unique_{j+1}|T_j=T ] > Pr[None_{j+1}|T_j=T] 
+o(\frac{1}{N^{1/4}})$ implies that
the probability of the second case is less than 
the probability of the first case
plus a term of order $\frac{1}{N^{1/4}}$ times the number of repetitions
for the while loop. The number of repetitions is $O(k \log N)$,
because $|T_{j+1}|\leq \frac{2k}{2k+1} (1+\frac{1}{2k^2}) |T_j|
\leq (1-\frac{1}{5k})|T_j|$.
Therefore, the probability of the second case is less than 
the probability of the first case
plus a term of order $o(\frac{k \log N}{N^{1/4}})=o(1)$.

It remains to prove the lemma.

\proof [of Lemma \ref{lem:sampling}]
We fix the permutations $\pi_1$, $\ldots$, $\pi_{j-1}$
and let $\pi_j$ be chosen uniformly
at random from the family of permutations given by
Theorem \ref{thm:kwise2}.

We consider two cases. The first case is when $T_j$ contains
many $k$-collisions. We show that, in this case, 
the lemma is true because the probability of $None_{j+1}$ is 
small (of order $o(\frac{1}{N^{1/4}})$). The second case
is if $T_j$ contains few $k$-collisions. In this case, we
pick one $x$ such that there are at least $k$ elements $i$, $x_i=x$.
We compare the probabilities that
\begin{itemize}
\item
$T_{j+1}$ contains no $k$-collisions;
\item
$T_{j+1}$ contains exactly one $k$-collision, consisting of $i$ with $x_i=x$.
\end{itemize}
The first event is the same as $None_{j+1}$, the second event
implies $Unique_{j+1}$. We prove the lemma by showing 
that the probability of the second event is at least the probability of
the first event minus a small amount. This is proven by
first conditioning on $T_{j+1}$ containing no $k$-collisions
consisting of $i$ with $x_i \neq x$ and then comparing
the probability that less than $k$ of $i:x_i=x$ belong to $T_{j+1}$
with the probability that exactly $k$ of $i:x_i=x$ belong to $T_{j+1}$.

\noindent
{\bf Case 1.}
$T_j$ contains at least $\log N$ pairwise disjoint sets 
$S_l=\{i_{l, 1}, \ldots, i_{l,k}\}$
with $x_{i_{l, 1}}=\ldots=x_{i_{l, k}}$.

Let $S=S_1\cup S_2\ldots \cup S_{\log N}$.
If event $None_{j+1}$ occurs, at least $\log N$ of 
$\pi_j \pi_{j-1} \ldots \pi_1 (i)$, $i\in S$
(at least one from each of sets $S_1$, $\ldots$, $S_{\log N}$)
must belong to $\{\lceil \frac{2k}{2k+1} q_j \rceil+1, \ldots, q_j\}$.
By the next claim, this probability is almost the
same as the probability that at least $\log N$ 
of $k\log N$ random elements of $[q_j]$ belong
to $\{\lceil \frac{2k}{2k+1} q_j \rceil+1, \ldots, q_j\}$.


\begin{Claim}
\label{claim:uniform}
Let $S\subseteq T_j$, $|S|\leq 2k\log N$.
Let $V\subseteq [q_j]^{|S|}$. Let $p$ be the probability
that $(\pi_j \pi_{j-1} \ldots \pi_1 (i))_{i\in S}$ belongs to
$V$ and let $p'$ be the probability that a tuple consisting
of $|S|$ uniformly random elements of $[q_j]$ belongs to $V$.
Then,
\[ |p-p'| \leq \frac{|S|^2+1}{q_j} .\]  
\end{Claim}

\proof
Let $S'=\{ \pi_{j-1}\ldots\pi_1 (i) | i\in S\}$.
Then, $p$ is the probability that 
$(\pi_j(i))_{i\in S'}$ belongs to $V$.
Let $p''$ be the probability that $(v_1, \ldots, v_{|S|})$
belongs to $V$, for $(v_1, \ldots, v_{|S|})$ picked uniformly
at random among all tuples of $|S|$ distinct elements of $[q_j]$.
By Definition \ref{def:approx-k}, $|p-p''|\leq \frac{1}{N}$.

It remains to bound $|p''-p'|$. If $(v_1, \ldots, v_{|S|})$ is picked 
uniformly at random among tuples of distinct elements, every tuple 
of $|S|$ distinct elements has a probability 
$\frac{1}{q_j(q_j-1)\ldots(q_j-|S|+1)}$ and the tuples of non-distinct
elements have probability 0. If $(v_1, \ldots, v_{|S|})$ is 
uniformly at random among all tuples, every tuple has probability
$\frac{1}{q_j^{|S|}}$. Therefore,
\[ \frac{q_j(q_j-1)\ldots(q_j-|S|+1)}{q_j^{|S|}} p'' \leq p' \leq
\frac{q_j\ldots(q_j-|S|+1)}{q_j^{|S|}} p'' + \left(1-
\frac{q_j\ldots(q_j-|S|+1)}{q_j^{|S|}} \right) ,\]
which implies 
\[ |p'-p''| \leq 1-\frac{q_j(q_j-1)\ldots(q_j-|S|+1)}{q_j^{|S|}} .\]
We have
\[ 1-\frac{q_j(q_j-1) \ldots(q_j-|S|+1)}{q_j^{|S|}} \leq
1- \left(\frac{q_j-|S|}{q_j}\right)^{|S|} \leq
1- \left(1-\frac{|S|^2}{q_j}\right) = \frac{|S|^2}{q_j} .\]
\qed

The probability that, out of $k\log N$ uniformly random 
$i_1, \ldots, i_{k\log N}\in\{1, \ldots, q_j\}$, 
at least $\log N$ belong to 
$\{\lceil \frac{2k}{2k+1} q_j \rceil+1, \ldots, q_j\}$
can be bounded using Chernoff bounds \cite{MU}.
Let $X_l$ be a random variable that is 1 if 
$i_l\in \{\lceil \frac{2k}{2k+1} q_j \rceil+1, \ldots, q_j\}$.
Let $X=X_1+\ldots+X_{k\log N}$.
We need to bound $Pr[X \geq \log N]$.
We have $E[X]=k \log N\cdot E[X_1]=\frac{k}{2k+1} \log N - o(1)$
and
\[ Pr[X\geq \log N]< \left(\frac{e^{(k+1)/(2k+1)}}{\frac{2k+1}{k}}
\right)^{\log N} = e^{-0.316..\log N} = o\left(\frac{1}{N^{1/4}}\right), \]
with the first inequality following from Theorem 4.4 of \cite{MU}
($Pr[X\geq (1+\delta) E[X]]<(\frac{e^{\delta}}{(1+\delta)^{1+\delta}})^{E[X]}$
for $X$ that is a sum of independent identically distributed 0-1
valued random variables).
\comment{

\begin{equation}
\label{eq-sum-ind} 
\sum_{m=\log N}^{k\log N} {k\log N \choose m} \left( \frac{1}{2k} \right)^m
\left( \frac{2k-1}{2k}\right)^{k\log N-m} .
\end{equation}
For $m\geq log N$, we have 
${k \log N \choose m+1}={k\log N \choose m}\frac{k\log N-m}{m+1}
\leq k {k\log N \choose m}$.
Therefore,
the sum (\ref{eq-sum-ind}) is at most $k\log N-m+1\leq k\log N$ times
\[ {k\log N \choose \log N} \left( \frac{1}{2k+1} \right)^{\log N}
\left( \frac{2k}{2k+1}\right)^{k\log N-\log N} \leq 
\left(\frac{e k \log N}{\log N}\right)^{\log N} \left( 
\frac{1}{2k+1} \right)^{\log N}
\left( \frac{2k}{2k+1}\right)^{k\log N-\log N} \]
\[ < \left( \frac{e}{2} \left(\frac{2k}{2k+1}\right)^{k-1} \right)^{\log N} =
\left(\frac{e}{2} \frac{1-o(1)}{\sqrt{e}} \right)^{\log N} =
\left(\frac{1-o(1)\sqrt{e}}{2} \right)^{\log N} .\]
Since $\frac{\sqrt{e}}{2}<\frac{1}{\sqrt[4]{2}}$, this means that the
sum (\ref{eq-sum-ind}) is $o(2^{-\log N/4})=o(\frac{1}{N^{1/4}})$.}
By combining this bound with Claim \ref{claim:uniform}, 
the probability of $None_{j+1}$ 
is 
\[ o\left(\frac{1}{N^{1/4}}\right)+
\frac{(k\log N)^2+1}{q_j} =
o\left(\frac{1}{N^{1/4}}\right) ,\]
where we used $q_j\geq |T_j|\geq \sqrt{N}$ (otherwise,
the algorithm finishes the while loop).

\noindent
{\bf Case 2.}
$T_j$ contains less than $\log N$ pairwise disjoint sets $S_l=\{i_{l, 1}, \ldots, i_{l,k}\}$
with $x_{i_{l, 1}}=\ldots=x_{i_{l, k}}$.

Let $S$ be the set of all $i$ such that $x_i$ is a
part of a $k$-collision among $x_i$, $i\in T_j$.

\begin{Claim}
$|S|<2 k\log N$.
\end{Claim}

\proof
We first select a maximal collection of pairwise
disjoint $S_l$. This collection contains less than $k\log N$ elements.
It remains to prove that $|S-\cup_l S_l|< k\log N$.

Since the collection $\{S_l\}$ is maximal, any $k$-collision between 
$x_i$, $i\in T_j$ must involve at least one element from $\cup_l S_l$.
Therefore, for any $x$, $S\setminus \cup_l S_l$ contains at most $k-1$ values $i$ 
with $x_i=x$. Also, there are less than $\log N$ possible $x$ 
because any $k$-collision must involve an element from 
one of sets $S_l$ and there are less than $\log N$ sets $S_l$. 
This means that  $|S-\cup_l S_l|< (k-1)\log N$.
\qed

Let $y_1, y_2, \ldots$ be an enumeration of all distinct $y$ such that
$T_j$ contains a $k$-collision $i_1, \ldots, i_k$ with
$x_{i_1}=\ldots=x_{i_k}=y$.
Let $UniqueColl_l$ be the event that $T_{j+1}$ contains exactly one
$k$-collision $i_1, \ldots, i_k$ with
$x_{i_1}=\ldots=x_{i_k}=y_l$ and $NoColl_l$ be the event
that $T_{j+1}$ contains no such collision.
The event $None_{j+1}$ is the same as $\bigwedge_l NoColl_l$.
The event $Unique_{j+1}$ is implied by 
$UniqueColl_1 \wedge \bigwedge_{l>1} NoColl_l$.
Therefore, it suffices to show
\begin{equation}
\label{eq-nocoll} Pr\left[\bigwedge_l NoColl_l\right] <
Pr\left[UniqueColl_1 \wedge \bigwedge_{l>1} NoColl_l\right] 
+ \frac{2((2k\log N)^2+1)}{q_j} .
\end{equation} 

The events $UniqueColl_l$ and $NoColl_l$ are equivalent to the cardinality of
\[ \left\{i:x_i=y_l, i\in T_j \mbox{ and } 
\pi_j\ldots\pi_1(i) \in \left\{1, \ldots, 
\left\lceil \frac{2k}{2k+1} q_j \right\rceil \right\} \right\} \]
being exactly $k$ and less than $k$, respectively.

By Claim \ref{claim:uniform}, the probabilities 
of both $\bigwedge_l NoColl_l$ and 
$UniqueColl_1 \wedge \bigwedge_{l>1} NoColl_l$ change by at most 
$\frac{(2k\log N)^2+1}{N}$ if we replace $(\pi_j \ldots \pi_1(i))_{i\in S}$
by a tuple of $|S|$ random elements of $[q_j]$.
Then, the events $NoColl_l$ and $UniqueColl_l$ are independent
of events $NoColl_{l'}$ and $UniqueColl_{l'}$ for $l'\neq l$. 
Therefore,
\[ Pr\left[\bigwedge_l NoColl_l\right] = Pr[NoColl_1] \prod_{l>1} Pr[NoColl_l] ,\]
\[ Pr\left[UniqueColl_1 \wedge \bigwedge_{l>1} NoColl_l\right] = 
Pr[UniqueColl_1] \prod_{l>1} Pr[NoColl_l] .\]
This means that, to show (\ref{eq-nocoll}) for the actual 
probability distribution $(\pi_j \ldots \pi_1(i))_{i\in S}$,
it suffices to prove $Pr[UniqueColl_1]\geq Pr[NoColl_1]$
for tuples consisting of $|S|$ random elements.

Let $I$ be the set of all $i\in T_j$ such that $x_i=y_1$. Let $m=|I|$.
Notice that $m\geq k$ (by definition of $x$ and $I$).
Let $P_l$ be the event that exactly $l$ of 
$\pi_j \ldots \pi_1 (i)$, $i\in I$ belong to $T_{j+1}$.
Then, $Pr[UniqueColl_1]=Pr[P_k]$ and 
$Pr[NoColl_1]=\sum_{l=0}^{k-1} Pr[P_l]$.
When $\pi_j \ldots \pi_1 (i)$, $i\in I$ are replaced 
by random elements of $[q_j]$, we have
\[ Pr[P_l] = {m \choose l} 
\left( 1-\frac{1}{2k+1} \right)^{l} \left( \frac{1}{2k+1} \right)^{m-l} ,\]
\[ \frac{Pr[P_{l}]}{Pr[P_{l+1}]} = 
\frac{{m \choose l}}{{m\choose l+1}} 
\cdot \frac{1}{2k+1} \cdot \frac{1}{1-\frac{1}{2k+1}} =
 \frac{l+1}{m-l} \cdot \frac{1}{2k} .\] 
For $l\leq k-1$, we have 
$\frac{l+1}{m-l} \frac{1}{2k}\leq k \frac{1}{2k}=\frac{1}{2}$.
This implies $Pr[P_l] \leq \frac{1}{2^{k-l}} Pr[P_k]$ and
\[ \sum_{l=0}^{k-1} Pr[P_l] \leq \left( 
\sum_{l=0}^{k-1} \frac{1}{2^{k-l}} \right) Pr[P_k] \leq Pr[P_k] \]
which is equivalent to $Pr[NoColl_1]\leq Pr[UniqueColl_1]$.
\comment{
Let $p$ be the probability of no $k$ equal elements 
from $x_j$, $T_i-I$ being in $T_{i+1}$. 
Event $None_{i+1}$ occurs if two conditions are satisfied:
\begin{itemize} 
\item at most $k-1$ elements from $J$ are in set $T_{i+1}$ and 
\item no $k$ equal elements from $x_j$, $j\in T_i-J$ are in $T_{i+1}$.
\end{itemize}
The two conditions are independent and the probability of the second condition.
Therefore, $Pr[None_{i+1}|T_i=T]= p \sum_{l=0}^{k-1} Pr[P_l] < p Pr[P_k]$. 

Consider the case when exactly $k$ elements from $J$ are in set $T_{i+1}$ and 
no $k$ equal elements from $x_j$, $j\in T_i-J$ are in $T_{i+1}$.
This happens with probability $p Pr[P_k]$.
Therefore, $Pr[Unique_{i+1}|T_i=T]\geq p Pr[P_k]$ and
$Pr[Unique_{i+1}|T_i=T]\geq Pr[None_{i+1}|T_i=T]$.
}
\qed

\section{Running time and other issues}
\label{sec:misc}

\subsection{Comparison model}

Our algorithm can be adapted to the model of comparison queries
similarly to the algorithm of \cite{Distinctness}. 
Instead of having the register $\otimes_{j\in S} \ket{x_j}$,
we have a register $\ket{j_1, j_2, \ldots, j_r}$
where $\ket{j_l}$ is the index of the $l^{\rm th}$ smallest
element in the set $S$.
Given such register and $y\in[N]$, we can add 
$y$ to $\ket{j_1, \ldots, j_r}$ by binary search 
which takes $O(\log N^{k/(k+1)})=O(\log N)$ queries.
We can also remove a given $x\in[N]$ in
$O(\log N)$ queries by reversing this process. 
This gives an algorithm with $O(N^{k/(k+1)}\log N)$ queries.

\subsection{Running time}

So far, we have shown that our algorithm solves element $k$-distinctness
with $O(N^{k/(k+1)})$ queries. 
In this section, we consider the actual running time of
our algorithm (when non-query transformations 
are taken into account).

{\bf Overview.}
All that we do between 
queries is Grover's diffusion operator which can be implemented in 
$O(\log N)$ quantum time and some data structure operations
on set $S$ (for example, insertions and deletions). 

We now show how to store 
$S$ in a classical data structure which supports the necessary
operations in $O(\log^4 (N+M))$ time.
In a sufficiently powerful quantum model, it is possible to transform 
these $O(\log^4 (N+M))$ time classical operations
into $O(\log^c (N+M))$ step quantum computation.
Then, our quantum algorithm runs in $O(N^{k/(k+1)}\log^c (N+M))$ steps.
We will first show this for the standard query model and
then describe how the implementation should be modified for
it to work in the comparison model.

\medskip
{\bf Required operations.}
To implement algorithm \ref{MainAlg}, we need the following operations:
\begin{enumerate}
\item
Adding $y$ to $S$ and storing $x_y$ (step \ref{step:add} of algorithm \ref{alg:qw});
\item
Removing $y$ from $S$ and erasing $x_y$ (step \ref{step:remove} of algorithm \ref{alg:qw});
\item
Checking if $S$ contains $i_1, \ldots, i_k$, $x_{i_1}=\ldots=x_{i_k}$
(to perform the conditional phase flip 
in step \ref{step1} of algorithm \ref{MainAlg});
\item
Diffusion transforms on $\ket{x}$ register in steps \ref{diffusion1} and
\ref{diffusion2} of algorithm \ref{alg:qw}.
\end{enumerate}

\medskip
{\bf Additional requirements.}
Making a data structure part of quantum algorithm creates two
subtle issues. 
First, there is the uniqueness problem.
In many classical data structures, the same set $S$ can be
stored in many equivalent ways, depending on the order in which elements
were added and removed. In the quantum case, this would mean that the basis
state $\ket{S}$ is replaced by many states $\ket{S^1}$,
$\ket{S^2}$, $\ldots$ which in addition to $S$ store some information
about the previous sets. This can have a very bad result. 
In the original quantum algorithm, we might have
$\alpha\ket{S}$ interfering with $-\alpha\ket{S}$, resulting
in 0 amplitude for $\ket{S}$. If $\alpha\ket{S}-\alpha\ket{S}$
becomes $\alpha\ket{S^1}-\alpha\ket{S^2}$, there is no
interference between $\ket{S^1}$ and $\ket{S^2}$ and the result
of the algorithm will be different.

To avoid this problem, we need a data structure where the same set
$S\subseteq [N]$ is always stored in the same way,
independent of how $S$ was created. 

Second, if we use a classical subroutine, it must terminate in a
fixed time $t$. Only then, we can replace it by an $O(poly(t))$ time
quantum algorithm. The subroutines that take time $t$ on average 
(but might take longer time sometimes) are not acceptable. 

\medskip
{\bf Model.}
To implement our algorithm, we use standard quantum circuit
model, augmented with gates for random access to a quantum
memory. A random access gate takes three inputs: $\ket{i}$, 
$\ket{b}$ and $\ket{z}$, with $b$ being a single qubit,
$z$ being an $m$-qubit register and $i\in[m]$.
It then implements the mapping
\[ \ket{i, b, z}\rightarrow \ket{i, z_i, z_1 \ldots z_{i-1} b z_{i+1} \ldots z_m} .\]
Random access gates are not commonly used in quantum algorithms
but are necessary in our case because, otherwise, simple data 
structure operations (for example, removing $y$ from $S$) which
require $O(\log N)$ time classically would require 
$\Omega(r)$ time quantumly.

In addition to random access gates, we allow 
the standard one and two qubit gates \cite{Barenco}.

\medskip
{\bf Data structure:overview.}
Our data structure is a combination of a hash table and a skip list.
We use the hash table to store pairs $(i, x_i)$ in the memory
and to access them when we need to find $x_i$ for a given $i$.
We use the skip list to keep the items sorted in the order of
increasing $x_i$ so that, when a new element $i$ is added to $S$,
we can quickly check if $x_i$ is equal to any of $x_j$, $j\in S$. 

We also maintain a variable $v$ counting the number of
different $x\in[M]$ such that the set $S$ contains $i_1, \ldots, i_k$
with $x_{i_1}=\ldots=x_{i_k}=x$.

\medskip
{\bf Data structure:hash table.}
Our hash table consists of $r$ buckets,
each of which contains memory for $\lceil\log N\rceil$ entries.
Each entry uses $O(\log^2 N+\log M)$ qubits.
The total memory is, thus, $O(r\log^3 (N+M))$, slightly more
than in the case when we were only concerned about the number
of queries.

We hash $\{1, \ldots, N\}$ to the $r$ buckets using 
a fixed hash function $h(i)=\lfloor i\cdot r/N \rfloor+1$.
The $j^{\rm th}$ bucket stores pairs $(i, x_i)$ for $i\in S$ such
that $h(i)=j$, in the order of increasing $i$. 

In the case if there are more than $\lceil\log N\rceil$ entries
with $h(i)=j$, the bucket only stores $\lceil \log N\rceil$
of them. This means that our data structure misfunctions.
We will show that the probability of that happening is small.

Besides the $\lceil\log N\rceil$ entries, each bucket also 
contains memory for storing $\lfloor\log r\rfloor$ counters
$d_1, \ldots, d_{\lfloor \log r\rfloor}$.
The counter $d_1$ in the $j^{\rm th}$ bucket counts the number
of $i\in S$ such that $h(i)=j$. 
The counter $d_l$, $l>1$ is only used if $j$ is divisible by $2^l$.
Then, it counts the number of $i\in S$ such that 
$j-2^l+1 \leq h(i)\leq j$.

\begin{figure*}
\begin{center}
\epsfxsize=5.5in
\hspace{0in}
\epsfbox{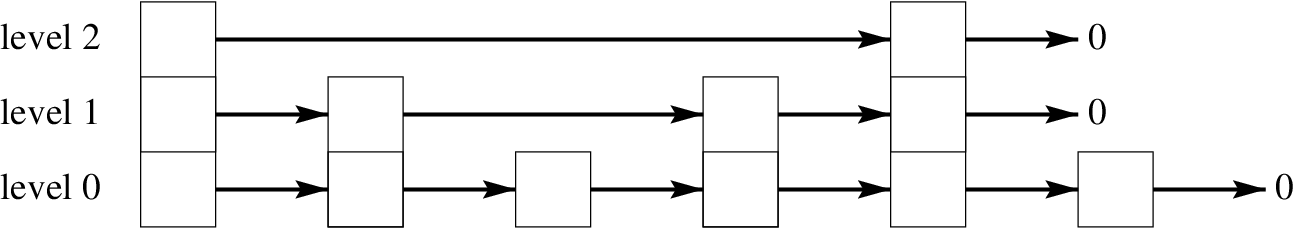}
\caption{A skip list with 3 levels}
\label{fig:skip}
\end{center}
\end{figure*}

The entry for $(i, x_i)$ contains $(i, x_i)$, together 
with a memory for $\lceil\log N\rceil+1$ pointers to other entries that
are used to set up a skip list (described below).

\medskip
{\bf Data structure:skip list.}
In a skip list \cite{Pugh}, each $i\in S$
has a randomly assigned level $l_i$ between 
0 and $l_{max}=\lceil\log N\rceil$.
The skip list consists of $l_{max}+1$ lists, from 
the level-0 list to the level-$l_{max}$ list.
The level-$l$ list contains all $i\in S$ with $l_i\geq l$. 
Each element of the level-$l$ level list has a level-$l$ 
pointer pointing to the next element of the level-$l$ list
(or 0 if there is no next element). 
The skip list also uses one additional ``start" entry. 
This entry does not store any $(i, x_i)$
but has $l_{max}+1$ pointers, with the
level-$l$ pointer pointing to the first element
of the level-$l$ list.
An example is shown in figure \ref{fig:skip}.

In our case, each list is in the order of increasing $x_i$.
(If several $i$ have the same $x_i$, they are ordered by $i$.)
Instead of storing an adress for a memory location, pointers store
the value of the next element $i\in S$. Given $i$, we 
can find the entry for $(i, x_i)$ by computing $h(i)$ and
searching the $h(i)^{\rm th}$ bucket.

Given $x$, we can search the skip list
as follows:
\begin{enumerate} 
\item 
Traverse the level-$l_{max}$ list until we find the last element
$i_{l_{max}}$ with $x_{i_{l_{max}}} < x$. 
\item 
For each $l=l_{max}-1, l_{max}-2, \ldots, 0$,
traverse the level-$l$ list, starting at
$i_{l+1}$, until the last element $i_{l}$ with $x_{i_l}< x$. %
\end{enumerate}
The result of the last stage is $i_0$, the last element of the level-0
list (which contains all $i\in S$) with $x_{i_0}<x$.
If we are given $i$ and $x_i$, a similar search can find the last
element $i_0$ which satisfies either $x_{i_0}<x_i$ or $x_{i_0}=x_i$
and $i_0<i$. This is the element which would precede $i$, if $i$
was inserted into the skip list. 


It remains to specify the levels $l_i$.
The level $l_i$ is assigned to each $i\in[N]$ before the beginning of 
the computation and does not change during the computation.
$l_i$ is equal to $j$ with probability $1/2^{j+1}$ for $j<l_{max}$
and probability $1/2^{l_{max}}$ for $j=l_{max}$.
\comment{For $j<l_{max}$, $i\in [N]$ is assigned to 
level $j$ with probability $1/2^{j+1}$. 
If $i\in [N]$ has not been assigned to 
any of levels $0, \ldots, l_{max}-1$, 
it gets assigned to level $l_{max}$. 
This happens with probability $1/2^{l_{max}}$.
}

The straightforward implementation (in which we chose the level
independently for each $i$) has the drawback that we have
to store the level for each of $N$ possible $i\in [N]$
which requires $\Omega(N)$ time to choose the levels and 
$\Omega(N)$ space to store them.
To avoid this problem, we define the levels 
using $l_{max}$ functions $h_1, h_2, \ldots, h_{l_{max}}:
[N]\rightarrow \{0, 1\}$.
$i\in[N]$ belongs to level $l$ (for $l< l_{max}$) if 
$h_1(i)=\ldots=h_{l}(i)=1$ but $h_{l+1}(i)=0$.
$i\in[N]$ belongs to level $l_{max}$ if 
$h_1(i)=\ldots=h_{l_{max}}(i)=1$.
Each hash function is picked uniformly at random from a 
$d$-wise independent family of hash functions (Theorem \ref{thm:kwise1}), 
for $d=\lceil 4 \log_2 N+1\rceil$.

In the quantum case, we augment 
the quantum state by an extra register holding 
$\ket{h_1, \ldots, h_{l_{max}}}$.
The register is initialized to a superposition in which every basis
state $\ket{h_1, \ldots, h_{l_{max}}}$ has an equal amplitude.
The register is then used to perform transformations dependent
on $h_1, \ldots, h_{l_{max}}$ on other registers. 

\comment{
For each $i\in S$, the entry for $(i, x_i)$ uses $l+1$ pointers
to other entries, where $l$ is the level of $i$. There is one pointer 
for each of levels $0, 1, \ldots, l$.  
We define $(i', x_{i'}) > (i, x_i)$ 
if either $x_{i'}>x_i$ or $x_i'=x_i$ and $i'>i$.
The level $l$ pointer is equal to $i_l$ 
such that $(i_l, x_{i_l})$ is minimal
subject to two constraints:
\begin{itemize}
\item
$i_l$ is of level at least $l$, and
\item
$(i_l, x_{i_l}) > (i, x_i)$. 
\end{itemize}
The level $l$ pointer is equal to 0 if there is no such $i_l$.

The skip list also uses one additional ``start" entry which does not
belong to any bucket. This entry does not store any $(i, x_i)$
but has $l_{max}+1$ pointers, with the
level-$l$ pointer equal to $i_l$ such that $i_l$ is of level at least
$l$ and $(i_l, x_{i_l})$ is minimal.
}

\medskip
{\bf Operations: insertion and deletion.}
To add $i$ to $S$, we first query the value $x_i$.
Then, we compute $h(i)$ and add $(i, x_i)$ to the 
$h(i)^{\rm th}$ bucket. If the bucket already contains
some entries, we may move some of them so that, after
inserting $(i, x_i)$, the entries are still in
the order of increasing $i$.
We then add 1 to the counter $d_1$ for the $h(i)^{\rm th}$ bucket
and the counter $d_l$ for the 
$(\lceil \frac{h(i)}{2^l}\rceil 2^l)^{\rm th}$
bucket, for each $l\in\{2, \ldots, \lfloor \log r\rfloor\}$.
We then update the skip list:

\begin{enumerate}
\item
Run the search for the last element before $i$ (as described earlier).
The search finds the last element $i_l$ before $i$ on
each level $l\in\{0, \ldots, l_{max}\}$.
\item
For each level $l\in\{0, \ldots, l_i\}$,
let $j_l$ be the level-$l$ pointer of $i_l$.
Set the level-$l$ pointer of $i$ to be equal to $j_l$
and the level-$l$ pointer of $i_l$ to be equal to $i$.
\comment{Set $l=l_{max}$; set $j$ to the ``start" entry;
\item
While $l\geq 0$, repeat:
\begin{enumerate}
\item
While level $l$ pointer of $j$ points to $j'$ 
such that $(j', x_{j'})<(i, x_i)$, let $j=j'$.
\item
Let $j'$ be the level $l$ pointer of $j$.
If the level of $i$ is at least $l$, set the level-$l$ pointer of $i$ equal 
to $j'$ and the level-$l$ pointer of $j$ equal to $i$.
\item 
Decrease $l$ by 1. 
\end{enumerate}}
\end{enumerate}

After the update is complete, we use the skip list to find
the smallest $j$ such that $x_{j}=x_i$ and then use level-0
pointers to count if the number of $j:x_{j}=x_i$ is less
than $k$, exactly $k$ or more than $k$.
If there are exactly $k$ such $j$, we increase $v$ by 1.
(In this case, before adding $i$ to $S$, there were $k-1$ such $j$
and, after adding $i$, there are $k$ such $j$.
Thus, the number of $x$ such that $S$ contains $i_1, \ldots, i_k$
with $x_{i_1}=\ldots=x_{i_k}=x$ has increased by 1.) 

An element $i$ can be deleted from $S$ by running this
procedure in reverse.

{\bf Operations: checking for $k$-collisions.}
To check for $k$-collisions in set $S$, we just check if $v>0$. 

\medskip
{\bf Operations: diffusion transform.}
As shown by Grover\cite{Grover}, the following transformation 
on $\ket{1}$, $\ldots$, $\ket{n}$ can be implemented with 
$O(\log n)$ elementary gates:
\begin{equation}
\label{eq-grover}
\ket{i}\rightarrow \left( -1+\frac{2}{n} \right) \ket{i}+
\sum_{i'\in [n], i'\neq i} \frac{2}{n} \ket{i'} .
\end{equation}
To implement our transformation in the step \ref{diffusion2}
of Algorithm \ref{alg:qw}, we need to implement 
a 1-1 mapping $f$ between between $S$ and $\{1, \ldots, |S|\}$.
Once we have such mapping, 
we can carry out the transformation
$\ket{y}\rightarrow \ket{f(y)}$ by
$\ket{y}\ket{0}\rightarrow \ket{y}\ket{f(y)} \rightarrow \ket{0}\ket{f(y)}$
where the first step is a calculation of $f(y)$ from $y$ and
the second step is the reverse of a calculation of $y$ from $f(y)$.
Then, we perform the transformation (\ref{eq-grover})
on $\ket{1}$, $\ldots$, $\ket{|S|}$ and then apply the transformation
$\ket{f(y)} \rightarrow \ket{y}$, mapping $\{1, \ldots, |S|\}$ back
to $S$.

The mapping $f$ can be defined as follows.
$f(y)=f_1(y)+f_2(y)$ where $f_1(y)$ is the number of items
$i\in S$ that are mapped to buckets $j$, $j<h(y)$ and
$f_2(y)$ is the number of items $y' \leq y$ that are mapped to bucket $h(y)$.
It is easy to see that $f$ is 1-1 mapping from $S$ to $\{1, \ldots, |S|\}$.
$f_2(y)$ can be computed by counting the number of items in bucket $h(y)$
in time $O(\log N)$.
$f_1(y)$ can be computed as follows:
\begin{enumerate}
\item
Let $i=0$, $l=\lfloor \log r\rfloor$, $s=0$.
\item
While $l\geq 0$ repeat:
\begin{enumerate}
\item
If $i+2^l<y$, add $d_l$ from the $(i+2^l)^{\rm th}$ bucket to $s$; let $i=i+2^l$;
\item
Let $l=l-1$;
\end{enumerate}
\item
Return $s$ as $f_1(y)$;
\end{enumerate}

The transformation in step \ref{diffusion1} of algorithm
\ref{alg:qw} is implemented, using a similar 1-1 mapping 
$f$ between between $[N]\setminus S$ and 
$\{1, \ldots, N-|S|\}$.

\medskip
{\bf Uniqueness.}
It is easy to see that a set $S$ is always stored in the same way.
The values $i\in S$ are always hashed to buckets by $h$ in the same
way and, in each bucket, the entries are located in the order
of increasing $i$. The counters counting the number of entries
in the buckets are uniquely determined by $S$.
The structure of the skip list is also uniquely determined,
once the functions $h_1, \ldots, h_{l_{max}}$ are fixed.

\medskip
{\bf Guaranteed running time.}
We show that,
for any $S$, the probability that lookup, insertion or deletion of some
element takes more than $O(\log^4 (N+M))$ steps is very small.
We then modify the algorithms for lookup, insertion or deletion so
that they abort after $c\log^4 (N+M)$ steps and show that this has 
no significant effect on the entire quantum search algorithm.
More precisely, let 
\[ \ket{\psi_t} = \sum_{S, y, h_1, \ldots, h_{l_{max}}} 
\alpha^t_{S, y} \ket{\psi_{S, h_1, \ldots, h_{l_{max}}}}
\ket{y}\ket{h_1, \ldots, h_{l_{max}}}   \]
be the state of the quantum algorithm after $t$ steps (each step being 
the quantum translation of one data structure operation), using 
quantum translations of the perfect data structure operations 
(which do not fail but may take more than $c\log^4 N$ steps).
Here,
$\ket{\psi_{S, h_1, \ldots, h_{l_{max}}}}$ stands for
the basis state corresponding to our data structure storing
$S$ and $x_i$, $i\in S$, using the hash functions 
$h_1, \ldots, h_{l_{max}}$. (Notice that the amplitude $\alpha^i_{S, y}$
is independent of $h_1, \ldots, h_{l_{max}}$, since 
$h_1, \ldots, h_{l_{max}}$ all are equally likely.)

We decompose $\ket{\psi_t}=\ket{\psi^{good}_t}+\ket{\psi^{bad}_t}$, 
with $\ket{\psi^{good}_t}$ consisting of $(S, h_1, \ldots, h_{l_{max}})$ 
for which the next operation successfully completes in $c\log^4 (N+M)$ steps
and $\ket{\psi^{bad}_t}$ consisting of $(S, h_1, \ldots, h_{l_{max}})$ 
for which the next operation fails to complete in $c\log^4 (N+M)$ steps.
Let $\ket{\psi'_t}$ be the state of the quantum algorithm
after $t$ steps using the imperfect data structure 
algorithms which may abort.
The next lemma is an adaptation of ``hybrid argument" by
Bennett et al. \cite{BBBV} to our context.

\begin{Lemma}
\label{lem:imperfect1}
\[ \|\psi_t-\psi'_t\| \leq \sum_{t'=1}^t 2\|\psi_{t'}^{bad}\| .\]
\end{Lemma}

\proof
By induction.
It suffices to show that 
\[ \|\psi_t-\psi'_t\|\leq \|\psi_{t-1}-\psi'_{t-1}\|+2\|\psi^{bad}_t\|. \]

To show that, we introduce an intermediate state
$\ket{\psi''_t}$ which is obtained by applying the perfect transformations
in the first $t-1$ steps and the transformation which may fail in
the last step. Then, 
\[ \| \psi_t-\psi'_t\| \leq \| \psi_t-\psi''_t\| + \| \psi''_t-\psi'_t\| .\]
The second term, $\| \psi''_t-\psi'_t\|$ is the same as
$\| \psi_{t-1}-\psi'_{t-1} \|$ because the states 
$\ket{\psi''_t}$ and $\ket{\psi'_t}$
are obtained by applying the same unitary transformation (quantum 
translation of a data structure transformation which may fail) to states
$\ket{\psi_{t-1}}$ and $\ket{\psi'_{t-1}}$, respectively.
To bound the first term, $\|\psi_t-\psi''_t\|$, let $U_p$ and $U_i$
be the unitary transformations corresponding to
perfect and imperfect version of the $t^{\rm th}$ data structure operation.
Then, $\ket{\psi_t}=U_p \ket{\psi_{t-1}}$ and
$\ket{\psi'_t}=U_i \ket{\psi_{t-1}}$. 
Since $U_p$ and $U_i$ only differ for $(S, h_1, \ldots, h_{l_{max}})$ 
for which the data structure operation does not finish in 
$c \log^4 N$ steps, we have
\[ \|\psi_t-\psi'_t\|= \|U_p \ket{\psi_{t-1}} - U_i \ket{\psi_{t-1}} \|=
\|U_p \ket{\psi^{bad}_{t-1}} - U_i \ket{\psi^{bad}_{t-1}} \|\leq
2\|\psi^{bad}_{t-1}\| .\]
\qed

\begin{Lemma}
\label{lem:imperfect2}
For every $t$, $\|\psi^{bad}_t\|=O(\frac{1}{N^{1.5}})$.
\end{Lemma}

\proof
We assume that there is exactly one $k$-collision 
$x_{i_1}=\ldots=x_{i_k}$. (If there is no $k$-collisions,
the checking step at the end of algorithm \ref{MainAlg}
ensures that the answer is correct. The case with  
more than one $k$-collision reduces to the case with exactly
one $k$-collision because of the analysis in section
\ref{sec:multiple}.)

By Lemma \ref{lem:symmetry}, every basis state $\ket{S,x}$
of the same type has equal amplitude. Also, all $h_1, \ldots, h_{l_{max}}$
have equal probabilities. Therefore, it suffices to show
that, for any fixed $s=|S\cap \{i_1, \ldots, i_k\}|$ and 
$t=|\{x\} \cap\{i_1, \ldots, i_k\}|$, the fraction of
$\ket{S, x, h_1, \ldots, h_{l_{max}}}$ for which the operation fails
is at most $\frac{1}{N^3}$.

There are two parts of the update operation which can fail:
\begin{enumerate}
\item
Hash table can overflow if more than $\lceil\log N\rceil$ elements $i\in S$ 
have the same $h(i)=h$;
\item
Update or lookup in the skip list can take more than $c\log^4 N$ steps.
\end{enumerate}

For the first part, let $s=|S\cap \{i_1, \ldots, i_k\}|$.
If more than $\lceil\log N\rceil$ elements $i\in S$
have $h(i)=j$, then at least $\lceil \log N\rceil-s$ of them
must belong to $[N]\setminus\{i_1, \ldots, i_k\}$.
We now show that, for a random
set $S\subseteq [N]\setminus \{i_1, \ldots, i_k\}$, $|S|=r-s$
the probability that more than $\lceil \log N \rceil-s$ of $i\in S$ 
satisfy $h(i)=j$ is small. 

We introduce random variables $X_1, \ldots, X_{r-s}$
with $X_l=1$ if $h$ maps the $l^{\rm th}$ element of $S$ to $j$.
We need to bound $X=X_1+\ldots+X_{r-s}$.
We have $\frac{N/r-s}{N-k} \leq E[X_l] \leq \frac{N/r}{N-k}$,
which means that $E[X_l]=\frac{1}{r}+O(\frac{1}{N})$.
(Here, we are assuming that $k$ is a constant. $s$ is also
a constant because $s\leq k$.)
Therefore, $E[X] = (r-s) E[X_l] = 1+o(1)$.

The random variables $X_l$ are negatively correlated: if 
one or more of $X_l$ is equal to 1, then the probability
that other variables $X_{l'}$ are equal to 1 decreases.
Therefore \cite{PS}, we can apply Chernoff bounds to
bound $Pr[X> \log N -s]$.
By using the bound $Pr[X\geq (1+\delta) E[X]]
<(\frac{e^{\delta}}{(1+\delta)^{1+\delta}})^{E[X]}$ \cite{MU,PS},
we get 
\[ Pr[X> \log N-s] < \frac{e^{\log N-s-1}}{(\log N-s)^{\log N-s}} =
o\left(\frac{1}{N^4}\right) .\]
\comment{
This probability is at most
\begin{equation}
\label{eq-hash}
 r \frac{{N/r-s \choose \lceil\log N\rceil -s}{N-\lceil \log N\rceil 
\choose r-\lceil\log N\rceil}}{{N-k \choose r-s}} ,
\end{equation}
with $r$ being the number of ways to choose $h$, 
${N/r-s \choose \lceil\log N\rceil -s}$ being the number of
ways to choose the $\lceil\log N\rceil -s$ elements 
$i\in S\setminus\{i_1, \ldots, i_k\}$ with $h(i)=j$
and ${N-\lceil \log N\rceil \choose r-\lceil\log N\rceil}$ being 
an upper bound on the
number of ways to choose the rest of $S$.
The denominator, ${N-k \choose r-s}$ is just the number of all 
ways to choose $S$.

The equation (\ref{eq-hash}) is at most
\[ r \left( \frac{e N}{r (\lceil\log N\rceil-s)} \right)^{\lceil\log N\rceil-s} 
\frac{(r-s) \ldots (r-c\log N+1)}{(N-s) \ldots (N-c\log N+1)} < \]
\[ r \left( \frac{e N}{r (\lceil\log N\rceil-s)} 
\frac{r}{N} \right)^{\lceil\log N\rceil-s} \leq
r \left( \frac{e}{\log N-s} \right)^{\log N-s} 
\leq N 2^{-(c-o(1))\log N \log \log N} = o(\frac{1}{N^{-4}}).\]

}

For the second part, we consider the time required for
insertion of a new element. (Removing an element requires the same
time, because it is done by running the insertion algorithm
in reverse.) Adding $(i, x_i)$ to the $(h(i))^{\rm th}$ bucket 
requires comparing $i$ to entries already in the bucket
and, possibly, moving some of the entries so that they remain
sorted in the order of increasing $i$. Since a bucket contains
$O(\log N)$ entries and each entry uses $\log^2 (N+M)$ bits,
this can be done in $O(\log^3 (N+M))$ time. 
Updating counters $d_l$ requires $O(\log N)$ time, for
each of $O(\log r)=O(\log N)$ counters.

To update the skip list, we first need to compute $h_1(i)$,
$\ldots$, $h_{l_{max}}(i)$. This is the most time-consuming step,
requiring $O(d\log^2 N)=O(\log^3 N)$ steps for each of 
$l_{max}=\lceil\log N\rceil$ functions $h_l$.
The total time for this step is $O(\log^4 N)$.
We then need to update the pointers in the skip list.
We show that, for any fixed $S, y$ (and random 
$h_1, \ldots, h_{l_{max}}$), the probability that updating 
the pointers in the skip list takes more than $c\log^4 N$ steps,
is small.

Each time when we access
a pointer in the skip list, it may take $O(\log^2 N)$ steps,
because a pointer stores the number $i$ of the next entry
and, to find the entry $(i, x_i)$ itself, we have to compute
$h(i)$ and search the $h(i)^{\rm th}$ bucket which may 
contain $\log N$ entries, each of which uses $\log N$ bits to store $i$. 
Therefore, it suffices to show that the probability of a 
skip list operation accessing more than
$c\log^2 N$ pointers is small.

We do that by proving that at most $d=4 \log N+1$ pointer accesses 
are needed on each of $\log N+1$ levels $l$. 
We first consider level 0. 
Let $j_1, j_2, \ldots$ be the elements of $S$ ordered
so that 
$x_{j_1}\leq x_{j_2}\leq x_{j_3} \ldots$
(and, if $x_{j_l}=x_{j_{l+1}}$ for some $j$, then $j_l<j_{l+1}$). 
If the algorithm requires more than $d$ pointer accesses on level 0,
it must be the case that, for some $i'$,
$j_{i'}$, $\ldots$, $j_{i'+d-1}$ are all at level 0.
That is equivalent to $h_1(j_{i'})=h_1(j_{i'+1})=\ldots=h_1(j_{i'+d-1})=0$. 
Since $h_1$ is $d$-wise independent, the probability
that $h_1(j_{i'})=\ldots=h_1(j_{i'+d-1})=0$ is $2^{-d}<N^{-4}$.

For level $l$ ($0<l<l_{max}$), we first fix the hash functions 
$h_1, \ldots, h_{l}$.
Let $j_1, j_2, \ldots$ be the elements of $S$ for which 
$h_1$, $\ldots$, $h_{l}$ are all 1, ordered so that 
$x_{j_1}\leq x_{j_2}\leq x_{j_3} \ldots$.
By the same argument, the probability that the algorithm
needs $d$ or more pointer accesses on level $l$ is 
the same as the probability that 
$h_{l+1}(j_{i'})=\ldots=h_{l+1}(j_{i'+d-1})=0$
for some $i'$ and this probability is at most $2^{-d}<N^{-4}$.
For level $l_{max}$, we fix hash functions 
$h_1, \ldots, h_{l_{max}-1}$
and notice that $i$ is on level $l_{max}$ 
whenever $h_{l_{max}}(i)=1$.
The rest of the argument is as before, with 
$h_{l_{max}}(j_{i'})=h_{l_{max}}(j_{i'+1})=\ldots=h_{l_{max}}(j_{i'+d-1})=1$
instead of $h_1(j_{i'})=h_1(j_{i'+1})=\ldots=h_1(j_{i'+d-1})=0$.

Since there are $\log N+1$ levels and $r$ elements of $S$,
the probability that the algorithm spends more than $k-1$
steps on one level for some element of $S$ is at most
$O(\frac{|S| \log N}{N^4})=O(\frac{1}{N^3})$. 

Therefore, $\|\psi^{bad}_t\|^2=O(\frac{1}{N^3})$ and
$\|\psi^{bad}_t\|=O(\frac{1}{N^{1.5}})$, proving the lemma. 
\qed

By Lemmas \ref{lem:imperfect1} and \ref{lem:imperfect2},
the distance between the final states of the ideal
algorithm (where the data structures never fail) and
the actual algorithm is of order 
$O(\frac{r}{N^{3/2}})=O(\frac{1}{N^{1/2}})$.
This also means that the probability distributions obtained
by measuring the two states differ by at most $O(\frac{1}{N^{1/2}})$,
in variational distance \cite{BV}. Therefore, the imperfectness
of the data structure operations does not have a significant effect.

{\bf Implementation in comparison model.}
The implementation in comparison model is similar, except
that the hash table only stores $i$ instead of $(i, x_i)$.

\section{Open problems}
\label{sec:open}

\begin{enumerate}
\item
{\bf Time-space tradeoffs.} 
Our optimal $O(N^{2/3})$-query algorithm 
requires space to store $O(N^{2/3})$ items.

How many queries do we need 
if algorithm's memory is restricted to $r$ items? 
Our algorithm needs $O(\frac{N}{\sqrt{r}})$ queries and this is the best
known. 
Curiously, the lower bound for deterministic algorithms 
in comparison query model\comment{\footnote{Up to our knowledge,
considering randomized algorithms or allowing queries of the form
considered in this paper gives no advantage but proving this has
been very difficult.}} is $\Omega(\frac{N^2}{r})$ 
queries \cite{Yao} which is quadratically more. 
This suggests that our algorithm might be optimal in this setting as well.
However, the only lower bound is the 
$\Omega(N^{2/3})$ lower bound for algorithms
with unrestricted memory \cite{Aaronson}. 
\item
\label{it:opt}
{\bf Optimality of $k$-distinctness algorithm.}
While element distinctness is known to require $\Omega(N^{2/3})$ queries,
it is open whether our $O(N^{k/(k+1)})$ query algorithm
for $k$-distinctness is optimal.

The best lower bound for $k$-distinctness is $\Omega(N^{2/3})$, by a following
argument. We take an instance of element distinctness $x_1, \ldots, x_N$ and transform
it into $k$-distinctness by repeating every element $k-1$ times.
If $x_1, \ldots, x_N$ are all distinct, there is no $k$ equal elements.
If there are $i, j$ such that $x_i=x_j$ among original $N$ elements,
then repeating each of them $k-1$ times creates $2k-2$ equal elements.
Therefore, solving $k$-distinctness on $(k-1)N$ elements requires at least
the same number of queries as solving distinctness on $N$ elements
(which requires $\Omega(N^{2/3})$ queries).
\comment{
\item
{\bf Generalizing our algorithm.}
It is plausible that the methods developed in this paper 
could be useful for other quantum algorithms as well.
To make them more applicable, it would be very useful to 
state our result in a general and easy-to-use form
(like amplitude amplification of \cite{BHMT} or quantum lower
bound theorems of \cite{AQLB} which can be easily used
without understanding every detail of their proofs).
A big step in this direction has been made by Szegedy \cite{Szegedy}.
}
\item
{\bf Quantum walks on other graphs.}
A quantum walk search algorithm based on similar ideas
can be used for Grover search on grids \cite{AKRS,CG}.
What other graphs can quantum-walks based algorithms
search? Is there a graph-theoretic property that determines
if quantum walk algorithms work well on this graph?

\cite{AKRS} and \cite{Szegedy} have shown that, for a class
of graphs, the performance of quantum walk depends on certain 
expressions consisting of graph's eigenvalues. In particular,
if a graph has a large eigenvalue gap, quantum walk search performs
well \cite{Szegedy}. A large eigenvalue gap is, however, 
not necessary, as shown by quantum search algorithms 
for grids \cite{AKRS,Szegedy}.
\end{enumerate}

{\bf Acknowledgments.}
Thanks to Scott Aaronson for showing that $k$-distinctness 
is at least as hard as 
distinctness (remark \ref{it:opt} in section \ref{sec:open}),
to Robert Beals, Greg Kuperberg and Samuel Kutin for 
pointing out the ``uniqueness''
problem in section \ref{sec:misc} 
and to Boaz Barak, Andrew Childs, Tung Chou,
Daniel Gottesman, Julia Kempe, Samuel Kutin, 
Frederic Magniez, Oded Regev, Mario Szegedy,
Tathagat Tulsi and anonymous referees for comments and discussions.

\end{document}